%% file: RBChain-main.tex
\documentclass[sigconf]{acmart}

\usepackage{amsmath,amsfonts}
\usepackage{algpseudocode}
\usepackage[ruled]{algorithm2e} % For algorithms
\usepackage{graphicx}
\usepackage{textcomp}
\usepackage{hyperref}
\usepackage{xcolor}
\usepackage{cleveref}
\usepackage{verbatim}

\algnewcommand{\Given}[1]{%
  \State \textbf{Given:}
  \Statex \hspace*{\algorithmicindent}\parbox[t]{.8\linewidth}{\raggedright #1}
}

\newtheorem{theorem}{Theorem}

\setcopyright{none}

\renewcommand\footnotetextcopyrightpermission[1]{} % removes footnote with conference information in first column
\makeatletter
\let\@authorsaddresses\@empty
\def\runningfoot{\def\@runningfoot{}}
\def\firstfoot{\def\@firstfoot{}}
\makeatother

\newcommand{\mybar}[1]{\mkern 1.5mu\overline{\mkern-1.5mu#1\mkern-1.5mu}\mkern 1.5mu}

\begin{document}
% Title portion
\title[A Regulatory System for Blockchains]{A Regulatory System for Optimal Legal Transaction Throughput in Cryptocurrency Blockchains}
\acmConference{}{}{}

\author{Aditya Ahuja}
\affiliation{%
  \institution{Indian Institute of Technology \\ Delhi}
  \city{New Delhi}
  \country{India}}
\email{aditya.ahuja@cse.iitd.ac.in}

\author{Vinay J. Ribeiro}
\affiliation{%
  \institution{Indian Institute of Technology Bombay}
  \city{Mumbai}
  \country{India}}
\email{vinayr@iitb.ac.in}

\author{Ranjan Pal}
\affiliation{%
  \institution{University of Michigan}
  \city{Ann Arbor}
  \country{USA}}
\email{palr@umich.edu}

%  \author{Leana Golubchik}
%  \affiliation{%
%  \institution{University of Southern California}
%  \city{Los Angeles}
% \email{leana@usc.edu}

\begin{abstract}
\input{00_abstract.tex}
\end{abstract}

%     <concept>
%        <concept_id>10002978.10003006.10003013</concept_id>
%        <concept_desc>Security and privacy~Distributed systems security</concept_desc>
%        <concept_significance>500</concept_significance>
%        </concept>
%    <concept>
%        <concept_id>10010405.10003550</concept_id>
%        <concept_desc>Applied computing~Electronic commerce</concept_desc>
%        <concept_significance>500</concept_significance>
%        </concept>
%    <concept>
%        <concept_id>10010405.10003550.10003551</concept_id>
%        <concept_desc>Applied computing~Digital cash</concept_desc>
%        <concept_significance>500</concept_significance>
%        </concept>
%    <concept>
%        <concept_id>10010405.10003550.10003552</concept_id>
%        <concept_desc>Applied computing~E-commerce infrastructure</concept_desc>
%        <concept_significance>500</concept_significance>
%        </concept>
%    <concept>
%        <concept_id>10010405.10003550.10003557</concept_id>
%        <concept_desc>Applied computing~Secure online transactions</concept_desc>
%        <concept_significance>500</concept_significance>
%       </concept>
%    <concept>
%        <concept_id>10010405.10003550.10003554</concept_id>
%        <concept_desc>Applied computing~Electronic funds transfer</concept_desc>
%        <concept_significance>500</concept_significance>
%        </concept>
%  </ccs2012>



% DO NOT use this command unless you want to change
% the default behavior
% \authorsaddresses{Authors' addresses: G.~Zhou, Computer Science
%   Department, College of William and Mary, 104 Jameson Rd,
%   Williamsburg, PA 23185, US, \path{gzhou@wm.edu}; V.~B\'eranger,
%   Inria Paris-Rocquencourt, Rocquencourt, France; A.~Patel, Rajiv
%   Gandhi University, Rono-Hills, Doimukh, Arunachal Pradesh, India;
%   H.~Chan, Tsinghua University, 30 Shuangqing Rd, Haidian Qu, Beijing
%   Shi, China; T.~Yan, Eaton Innovation Center, Prague, Czech Republic;
%   T.~He, C.~Huang, J.~A.~Stankovic University of Virginia, School of
%   Engineering Charlottesville, VA 22903, USA; T. F. Abdelzaher,
%   (Current address) NASA Ames Research Center, Moffett Field,
%   California 94035.}


\maketitle

% \renewcommand{\shortauthors}{G. Zhou et al.}

\input{01_introduction.tex}
\input{02_case-silkroad.tex}
\input{03_reg-sysmodel.tex}
\input{04_reg-proto.tex}
\input{05_comp-analysis.tex}
\input{06_relatedwork.tex}

\input{07_conclusion.tex}

% \section*{Acknowledgment}
% \noindent We would like to thank the Quantitative Evaluation and Design (QED) group at the University of Southern California for their technical support in the evaluation of ZenCouncil.

% Bibliography

\input{08_rbc-ref.tex}
% Appendices
\appendix
\input{09a_thm-proofs.tex}

\end{document}

%% file: 00_abstract.tex
Permissionless blockchain consensus protocols have been designed primarily for defining decentralized economies for the commercial trade of assets, both virtual and physical, using cryptocurrencies. In most instances, the assets being traded are \emph{regulated}, which mandates that the legal right to their trade and their trade value are determined by the governmental regulator of the jurisdiction in which the trade occurs. Unfortunately, existing blockchains do not formally recognise proposal of legal cryptocurrency transactions, as part of the execution of their respective consensus protocols, resulting in rampant illegal activities in the associated crypto-economies. 
In this contribution, we motivate the need for regulated blockchain consensus protocols with a case study of the illegal, cryptocurrency based, Silk Road darknet market. We present a novel regulatory framework for blockchain  protocols, for ensuring legal transaction confirmation as part of the blockchain distributed consensus. As per our regulatory framework, we derive conditions under which legal transaction throughput supersedes throughput of  traditional transactions, which are, in the worst case, an indifferentiable mix of legal and illegal transactions. Finally, we show that with a small change to the standard blockchain consensus execution policy (appropriately introduced through regulation), the legal transaction throughput in the blockchain network can be maximized.

%% file: 01_introduction.tex
\section{Introduction}

Decentralized financial institutions are a novel, emerging economic infrastructure. These institutions are based predominantly on cryptocurrencies for the trade of assets, both physical and virtual. It has been established that the collective market capitalization of cryptocurrencies is over \$1 trillion (as of early 2021) \cite{cc-mcap}. This shows promise in the long term presence and viability of cryptocurrency based markets, in competition with (and possibly as a replacement of) federally administered centralized financial institutions. However, these cryptocurrency based markets survive solely on the correctness of the underlying computational principles, which are a basis of the efficacy of these economies. More specifically, in order to sustain these cryptocurrency based decentralized economies, blockchain consensus protocols serve as a technical foundation. \\
Existing blockchain protocols for cryptocurrencies address one of (or any combination of) the following system goals: \emph{speed, security and decentralization}. Unfortunately, these system goals are necessary but insufficient. Illegal activities propelled through the strategic use of blockchain based cryptocurrencies, is a serious problem staring at the face of many world governments today \cite{silk-law}. These illegal activities exploit the permissionless nature of the blockchain networks for illegal trade, to strategically defeat regulation by obfuscating the jurisdictions of the blockchain users through anonymity, making federal legal rules inapplicable \cite{chain-reg-anon}. In this contribution, we introduce a fourth pillar of correctness for ensuring confidence in blockchain based cryptocurrencies: \emph{legality}. More specifically, we address the problem of legal cryptocurrency based trade of \emph{regulated assets:} assets whose value and trade terms are overseen by the respective governments\footnote{Regulated assets include but are not limited to the following asset classes: cryptocurrencies (evidence of regulation: \cite{evd-ccreg}), alternate virtual assets (for example copyrighted digital content), physical commodities (for example gold, water, energy), real-estate, etc.}.

\subsection*{The Problem Setting}

For fear of legal scrutiny and persecution, law abiding cryptocurrency blockchain users may sign up and get authorization from their respective governments (for instance, through the issuance of the BitLicense in New York, USA \cite{bitlicense}) to commit solely to legal cryptocurrency based trade of regulated assets, and the respective regulator (for instance, the New York State Department of Financial Services in the case of BitLicense \cite{bitlicense}) may license these law abiding blockchain users to legally participate in the crypto-economy. However, this does not preclude illegal transactions going on chain. Remembering that the blockchain network is permissionless, there will always exist unregulated transactors and executors of the consensus protocol that propose and mine/validate dubious transactions having an indeterminate legal status, while faithfully following the consensus protocol. Consequently, there opens up a competition in dubious versus legal block proposal between unregulated and regulated consensus protocol executors (respectively), and that competition is dependent on how much consensus resource (for instance, mining power in proof-of-work blockchains and stake in proof-of-stake blockchains) in the blockchain network, do the regulated consensus protocol executors possess as a whole. This finally results as a problem for the federal regulatory body to strategically decide on how much consensus resource to license in the blockchain network, and what block proposal strategies to advise to the regulated consensus protocol executors, for reasonable guarantees on legal transaction throughput. \\
\noindent The resultant open questions that we address, are: \emph{(i) given a permissionless blockchain network for cryptocurrency based regulated asset trade, can there exist a framework where, without degrading the anonymity/privacy of the blockchain users, legal transactions can be clearly identified and confirmed in the blockchain network?; and (ii) can we derive conditions, as a function of the regulated consensus resource in the blockchain network, in which legal transaction throughput supersedes dubious transaction throughput?}

\subsection*{Our Research Contributions}

In this paper, we first motivate the need for designing regulatory frameworks for legal cryptocurrency based trade of regulated assets, through the study of the Silk Road darknet online market \cite{silk-drug,silk-econ,silk-law} (Section \ref{sec:casestudy}). We then make the following research contributions.
\begin{enumerate}
    % \item \textsf{(A Case Study)} We study the Silk Road darknet online market \cite{silk-drug,silk-econ} in the context of the illegal trades due to anonymous cryptocurrency transactions. We then argue how this case study gives motivation for designing regulatory frameworks for legal cryptocurrency based regulated asset trade in general \cite{silk-law} (Section \ref{sec:casestudy}).

    \item \textsf{(A Regulatory Framework)} We define a regulated blockchain consensus system for existing permissionless blockchain protocols that support cryptocurrencies (Sections \ref{sec:sysmodel} and \ref{sec:reg-proto}), where the goal of the consensus protocol executors is to maximize their rewards from participating in the consensus protocol, and the mandate of the regulator is to maximize the legal transaction throughput in the blockchain protocol. 
    
    \item \textsf{(Block Proposal Competition)} Consequent to our regulated consensus framework, there ensues a competition in block proposal between regulated and unregulated consensus executors. We formalize this competition through a two player stochastic game \cite{sg-solan,bmg,bmg-pf} (Section \ref{sec:competition}), and, we show that: 
    \begin{enumerate}
        \item \textsf{(Under Immediate Block Release \cite{bmg})} When the regulator licenses between $58\%$ and $100\%$ of the consensus resource, the unregulated executors can do no better than adding legal blocks at the end of the longest unconfirmed branch of legal blocks (Section \ref{subsec:sg-ir}), thereby maximizing the legal transaction throughput.
        \item \textsf{(Under Immediate Block Release \cite{bmg}, With An Oversight Compliance Fee \cite{bmg-pf})} When the regulator licenses between $50\%$ and $58\%$ of the consensus resource, given that the regulator incentivizes to build on the branch of legal blocks with a pay forward scheme (similar to \cite{bmg-pf}), the unregulated executors can do no better than adding legal blocks at the end of the longest unconfirmed branch of legal blocks (Section \ref{subsec:sg-ir-ocf}), again maximizing the legal transaction throughput.
        \item \textsf{(Under Strategic Block Release \cite{btc-sm1,bmg})} When the regulator licenses between $33\%$ and $50\%$ of the consensus resource, given that regulated executors strategically release a subset of the blocks notarized by them (similar to the selfish mining attacks \cite{btc-sm1,btc-osm}), and the unregulated executors faithfully follow the longest branch rule, the regulated executors can prune/orphan some notarized but unconfirmed dubious blocks to increase the legal transaction throughput, beyond their fair share of legal transaction throughput. However, when the regulator licenses between $0\%$ and $33\%$ of the consensus resource, the regulated consensus executors can do no better than building on the longest unconfirmed branch of dubious blocks (Section \ref{subsec:sg-sr}), resulting in legal transaction throughput proportional to their consensus resource.
    \end{enumerate}

    % \item \textsf{(Regulator induced Full-Decentralization \cite{imposs-fulldecen})} Given that full decentralization in permissionless blockchains is impossible without a trusted third party \cite{imposs-fulldecen}, we briefly outline a framework where the (trusted) regulators can seed randomness in the blockchain network for a fully decentralized and consequently fair crypto-economy (Section \ref{sec:reg-decen}).  
\end{enumerate}

\begin{figure}
	\centering
	\includegraphics[width=0.95\linewidth]{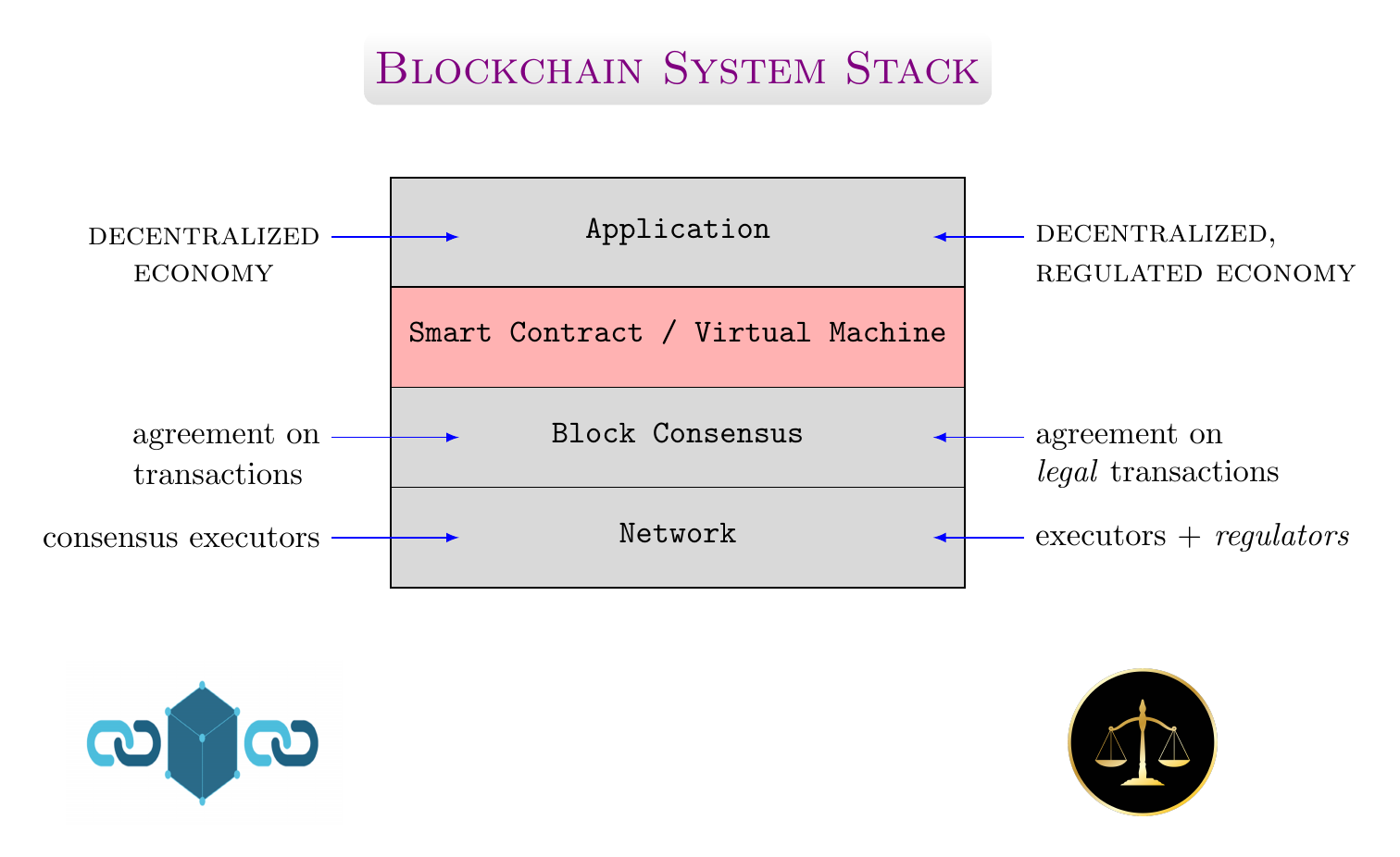}
	\caption{Traditional vs Regulated Blockchain Systems}
	\vspace{-1em}
	\label{fig:sys-stack}
\end{figure}

\noindent Blockchain based crypto-economies are defined using a four layer system stack \cite{sok-wang}. The difference between a traditional instantiation of the blockchain stack and an instantiation corresponding to our regulated blockchain system is depicted in Figure \ref{fig:sys-stack}.

% \draw{Traditional vs Regulated Blockchain System Stack}

%% file: 02_case-silkroad.tex
\section{Case Study: The Silk Road Dark Web Market}
\label{sec:casestudy}

\noindent We first give a brief study on the operation of the Silk Road Darknet Market, which flourished through the illegal use of the Bitcoin cryptocurrency. 

\subsection{An Overview of the Silk Road Marketplace}
\label{subsec:silk-overview}

% \aditya{``The most important institution of the Deep Web is anonymity. Each buyer and seller is known by a unique username; their true identity is secret. Users of the Deep Web, through forums and blogs, create a wealth of information to keep users updated on the happenings of the market (DarkNetMarkets, 2014). Black-market activity on the Deep Web is attractive because of the anonymity it provides. Cryptocurrencies such as Bitcoin (BTC) function like cash; they are untraceable. The TOR network anonymizes web traffic. PGP encryption programs mask data within emails sent between users. These three elements form the technological base upon which Deep Web black markets build, allowing exchange at a much lower cost than previously. Before this technology, sellers and buyers in the black market relied heavily on face-to-face interaction and building a reputation through personal encounters. This shift led to a flourishing peer-to-peer underground marketplace expanding on a global scale." \cite{silk-econ}} \\
\noindent Anonymity is the most prominent institution of the Deep Web. Each user (buyer and seller) is identified by a username, with a secret true identity. Deep Web users record market interactions through forums and blogs. Consequent to the anonymity it provides, black market activity over the Deep Web is highly feasible and attractive. Web traffic is anonymized through TOR, and Bitcoin serves as an untraceable virtual currency. Email interactions to discuss illicit transaction details are encrypted through PGP. These three network, email and currency elements serve as the technological foundation to build an illegal market, with low cost illegal transactions. Given that previously illegal transactors relied heavily on in-person deals and reputations built on personal encounters, Deep Web based illicit markets resulted in a paradigm shift for illegal activities on a global scale \cite{silk-econ}. \\
% \aditya{"The online drug marketplace called Silk Road has operated anonymously on the Deep Web since 2011. It is accessible through computer encrypting software (Tor) and is supported by online transactions using peer to peer anonymous and untraceable crypto-currency (Bit Coins). The study aimed to describe user motives and realities of accessing, navigating and purchasing on the Silk Road marketplace. Methods: Systematic online observations, monitoring of discussion threads on the site during four months of fieldwork and analysis of anonymous online interviews (n = 20) with a convenience sample of adult Silk Road users was conducted." \cite{silk-drug}}. \\
\noindent The Silk Road online (mostly) narcotics trade marketplace has flourished anonymously on the Deep Web since 2011. A study aimed at the discovering the realities and motives of operations (navigation and purchase) by drug users on the Silk Road marketplace has been conducted \cite{silk-drug}. The study was conducted through strategic online observations, four month long fieldwork on marketplace site discussion threads and twenty anonymous online interviews of a sample of adult users.
% ADD: \aditya{*Sections VI-A and VI-B from \cite{silk-law}.*}

\subsection{Illicit Transactions employing Bitcoin}
% \aditya{"A primary difference between traditional online sites, such as eBay, and the Silk Road is escrow implementation. Standard escrow requires the ability to undo a transaction. Fraudulent items are returned to the seller, and then the escrow service refunds the buyer. Hu et al. preface their model on the assumption that ‘in the case of fraud, [escrow] users lose only the service fee’ (Hu et al., 2004). Silk Road purchases cannot be undone; drug dealers do not provide return addresses. An escrow service cannot exist which simultaneously satisfies buyer and seller."\cite{silk-econ}.} \\
\noindent The Silk Road marketplace is inferior to traditional online sites (such as eBay), due to its bad Bitcoin \emph{escrow implementation}: the ability to undo a transaction. Standard escrow requires that if the trade is deemed fraudulent, the traded assets are returned to the seller, and the escrow service refunds the currency to the buyer. In case of fraud, market users only lose their service fee. Typical of illegal marketplaces, Silk Road purchases cannot be undone: drug sellers do not provide return addresses, and a perfect escrow service cannot exist (although a rudimentary escrow can) that satisfies both the buyer and the seller, simultaneously \cite{silk-econ}. \\
% \aditya{"Several participants observed that the process of accessing the site via the Tor browser, arranging credit with ‘Bit Coins’ and purchasing products on the site was time consuming and relatively difficult. Observational site data recorded that members have technical knowledge around internet security. ‘I obtain Bit Coins. I won’t go into the exact process for safety's sake, but basically I go into a bank with cash, fill out a slip, show absolutely no identifying information, and by the end of the day, I have Bit Coins in my Silk Road account. I then go to the vendor page, add the drug to my shopping cart, and input my encrypted address. Then I confirm the order and I wait. We also use an Escrow system so that vendors can’t scam you. So the Bit Coins aren’t directly delivered to them until I finalize my order, which I only do once the package arrives. Some vendors require early finalization, but I try not to deal with them’. (Participant 9, Male aged 20–25 years)." \cite{silk-drug}.} \\
\noindent A prominent drug study \cite{silk-drug} concluded that many narcotics buyers were not technically proficient and faced trouble in arranging for Bitcoin credit and accessing the Silk Road website via the Tor browser. An anonymous interview of an adult drug user revealed his drug procurement process over Silk Road. The said user was able to procure Bitcoins with minimal paperwork from a particular bank, with no self identifying information submitted to the bank. By the end of the day, the user had Bitcoins credited to his Silk Road account. (S)He then went to the narcotics' vendor webpage, added the drugs to his shopping cart and entered an encrypted postal address for the drug delivery, thereby confirming the order. The user even employed an escrow system to avoid being scammed by the vendor. Consequently, the Bitcoins are not delivered to the vendor until the drug package reached the user and the order was finalized.
% ADD: \aditya{Section VI-C from \cite{silk-law}.} \\

\noindent Thousands of Bitcoin worth approximately a billion US dollars connected with Silk Road based drugs and goods trade have been confiscated by the United States Justice Department, the biggest seizure in history of the agency \cite{silk-bust}. This motivates the need for preemptive monetary investment by the regulator to ensure only legal transactions are confirmed on-chain (as is suggested in Section \ref{subsec:sg-ir-ocf}) to minimize the cost associated with law enforcement and the recovery of illegal cryptocurrency, and reduce illegal activity through the blockchain. \\
\noindent Very recently, senior authorities from the US Treasury Department and the European Central Bank have also formally recognised the use of Bitcoin and other cryptocurrencies for illegal activities, and are strongly considering regulation in these digital economies \cite{btcnews-yellen,btcnews-lagarde}. \\

\noindent The drug study, seizures and recent concerns raised by authoritative figures in prominent financial bodies advocate that, at least the Bitcoin protocol, is rife with issues of confirmed illegal transactions, and appropriate regulatory mechanisms are need to be enforced to define legal, decentralized crypto-economies.  

\subsection{Motivation for a Regulated Blockchain Protocol}

Blockchain systems are defined with a four layer system stack (from bottom to top): the network layer, the consensus layer, the smart contract layer, and the application layer \cite{sok-wang}. Given the pressing need for a regulatory framework for blockchains, we now motivate how introducing regulation for blockchain transactors and consensus protocol executors is a prudent choice for the regulatory body.

\subsubsection*{Problems with Super-Consensus Layer Solutions}
% \aditya{Ambiguity: Difficult to tie $vk_i$ to $f_i$ at smart contract layer.} \\
% \aditya{Speed: A smart contract layer solution, as opposed to a consensus layer solution \cite{sok-wang}, would require time $t_{cons} + t_{SC} > t_{cons} \approx t_{reg-cons}$.}
Without consideration of the network and consensus layer, regulatory policy can only be enforced at the smart contract layer, given that the application layer is confined to the decentralized trade of regulated assets. Unfortunately, there are two problems associated with regulatory policy enforcement at the smart contract layer: (i) As it has been observed in Section \ref{subsec:silk-overview}, most times when a blockchain user is involved in illegal trade, its digital identity is difficult to map to a specific federal jurisdiction. This makes it hard to enforce regulation rules as a distributed program (which is the executed at the smart contract layer). (ii) A smart contract layer enforcement of regulation will take \emph{time to achieve consensus + time to execute the smart contract}. This would be slower in contrast to consensus layer only enforcement of regulation.

\subsubsection*{Regulating Blockchains at the Consensus Layer}
% \aditya{License transaction proposal and consensus resource to ensure legal transaction confirmation as part of blockchain consensus.}
One approach to regulate the blockchain system, to ensure legal transaction proposal and confirmation, is to license the blockchain consensus protocol executors who take it upon themselves to add (provably) legal transactions to their blocks, and run the consensus protocol on these blocks. We give details of this regulatory approach, next.

%% file: 03_reg-sysmodel.tex
\section{The Regulated Blockchain System Model}
\label{sec:sysmodel}

\noindent We now detail our regulated blockchain system model. We first give a short summary on the principles of blockchain consensus, and give some foundational definitions and assumptions. We then give the design goals and protocol features (with their motivations) for our regulatory framework. Finally, we give the threat model in our setting.

\subsection{A Brief on Blockchain Consensus}
The original consensus algorithm in a blockchain network is \emph{proof-of-work} (PoW), where the protocol participants, called \emph{miners}, solve a crypto-puzzle requiring a significant amount of computation, to add (notarize) blocks as part of the blockchain. On successfully adding a block to the blockchain, miners are eligible for a \emph{block reward} in the form of cryptocurrency tokens, for their effort. Given that computation is the only resource that dominates the result of consensus, miners might form a \emph{pool} of their computation power together \cite{btc-sm1} to increase their chances of solving the puzzle, adding their blocks to the blockchain, and consequently winning the block reward. \\
An alternate, dominant family of blockchains follow a \emph{proof-of-stake} (PoS) consensus regime. In proof-of-stake, blockchain protocol executors \emph{validate} the correctness of blocks as a function of the number of coins / digital tokens they possess in the blockchain network, which is referred to as their \emph{stake}. Remembering that digital tokens are a replicatable resource, one possible attack in these blockchain systems is that protocol executors can duplicate and pitch their coins in every block competing to be added to the blockchain, in order to maximize their chances of winning the block reward. This attack is called a \emph{nothing-at-stake} attack \cite{v-pos}.

\subsection{Stakeholders, Terminology, Assumptions, and Notation}

We define the stakeholders in our regulated decentralized economy, and give the associated terminology and assumptions, first.

\begin{itemize}
    \item \emph{Transaction and Block Types:} We will assume there are two types of transactions: \emph{legal} or \emph{dubious}. Transactions whose legal status can be established for certain will be called legal. All other transactions would be called dubious. A block that contains at least one dubious transaction will be called a \emph{dubious block}. A block that contains only legal transactions will be called a \emph{legal block}. A legal block generated by a regulated consensus protocol executor (defined later) will be called a \emph{regulated block}. 
    \item \emph{A Body of Regulators:} We will assume the existence of a cross-jurisdictional network of regulators, which are federal associations in-charge of ensuring legal and non-discriminatory practices in the institutions they oversee \cite{nn-regstate}. In our case the institution is the blockchain network. 
    \item \emph{Blockchain Transactors:} We will refer to blockchain users that define and propose cryptocurrency transactions as \emph{transactors}. The transactors can be of two types: regulated and unregulated. Regulated transactors will always propose transactions that are legal and verifiable within the blockchain network (\textsc{assumption:} The regulator will announce a license corresponding to each regulated transactor so that the blockchain network can verify the legality of transactions proposed by them). Unregulated transactors can propose either of legal and dubious transactions.
    \item \emph{Blockchain Executors:} We will refer to blockchain users that execute the consensus protocol as \emph{consensus executors}, or \emph{executors} for short. These executors will be miners in proof-of-work (PoW) blockchains, or validators in proof-of-stake (PoS) blockchains. We will assume that all executors are rational, and want to maximize their reward / revenue resulting from their participation in the consensus protocol (identical to the standard notion of rationality in \cite{bmg,bmg-pf}). We will also refer to the consensus resource associated with each executor, where the resource is hash power in PoW blockchains, or stake in PoS blockchains. Here too, we will assume executors will be of two types: regulated and unregulated. Regulated executors will propose blocks that contain only verifiably legal transactions, with an evidence of regulation of the said executor, and so their blocks will be called regulated blocks. Unregulated executors can propose blocks that may be either legal or dubious. When specifically dealing with PoS blockchains, we will consider Byzantine behaviour by unregulated validators through a nothing-at-stake attack \cite{v-pos}. Finally, we assume that unregulated executors cannot form selfish notarization pools, due to lack of trust in fair block reward distribution in the unregulated setting, given the pool administrator can be dishonest \cite{btc-minepractice}. However unregulated executors can coordinate amongst themselves for deciding individual notarization strategies that may maximize their individual expected block rewards.
    \item \emph{Block Notarization and Confirmation:} We say that a proposed block is \emph{notarized}\footnote{In the context of dubious blocks, the word `notarized' will bring out the violation in how unregulated executors validate/attest blocks with possibly illegal transactions.} (similar terminology in \cite{streamlet}), once it is successfully mined by a PoW executor, or successfully validated by a PoS executor (for instance, on receiving the threshold of validation votes in Algorand \cite{algorand}). We say that a block is \emph{confirmed} once it is sufficiently deep in the blockchain and all the transactions contained in it are confirmed/finalized (for instance, transactions in Bitcoin are confirmed once they are six blocks deep \cite{bitcoin}).
    \item \emph{Block Proposal in Discrete Time:} We assume that the consensus protocol executors have loosely synchronized clocks by employing protocols such as NTP \cite{algorand}. We will ignore network delays, and assume that blocks are proposed in discrete sequential epochs of time, with the time difference between two consecutive epochs equal to the block proposal time of the base protocol (for instance, $10$ minutes in Bitcoin \cite{bitcoin} and $22$ seconds in Algorand \cite{algorand}). Also, the epoch number of every block will also be equal to the block's height.
    \item \emph{Dubious and Legal Block Branches:} Given the block proposal competition ensuing between two categories (regulated and unregulated) of consensus executors, there would exist a fork in the blockchain with block branches corresponding to each category of executors. We would refer to the block where this fork originates as the root block (this would be the block at the end of the trunk of the unambiguous part of the blockchain). The block branch corresponding to regulated executors will be referred to as the \emph{legal branch} (as all blocks in this branch will be verifiably legal), and that corresponding to unregulated executors will be referred to as a \emph{dubious branch}. Given any one of the legal or dubious branches, we would refer to the most recently notarized block in a branch as a \emph{frontier} block for that branch, and every other notarized block as an \emph{interior} block for that branch.
\end{itemize}

% \draw{Table of Notation}
\begin{figure}
	\centering
	\includegraphics[width=0.95\linewidth]{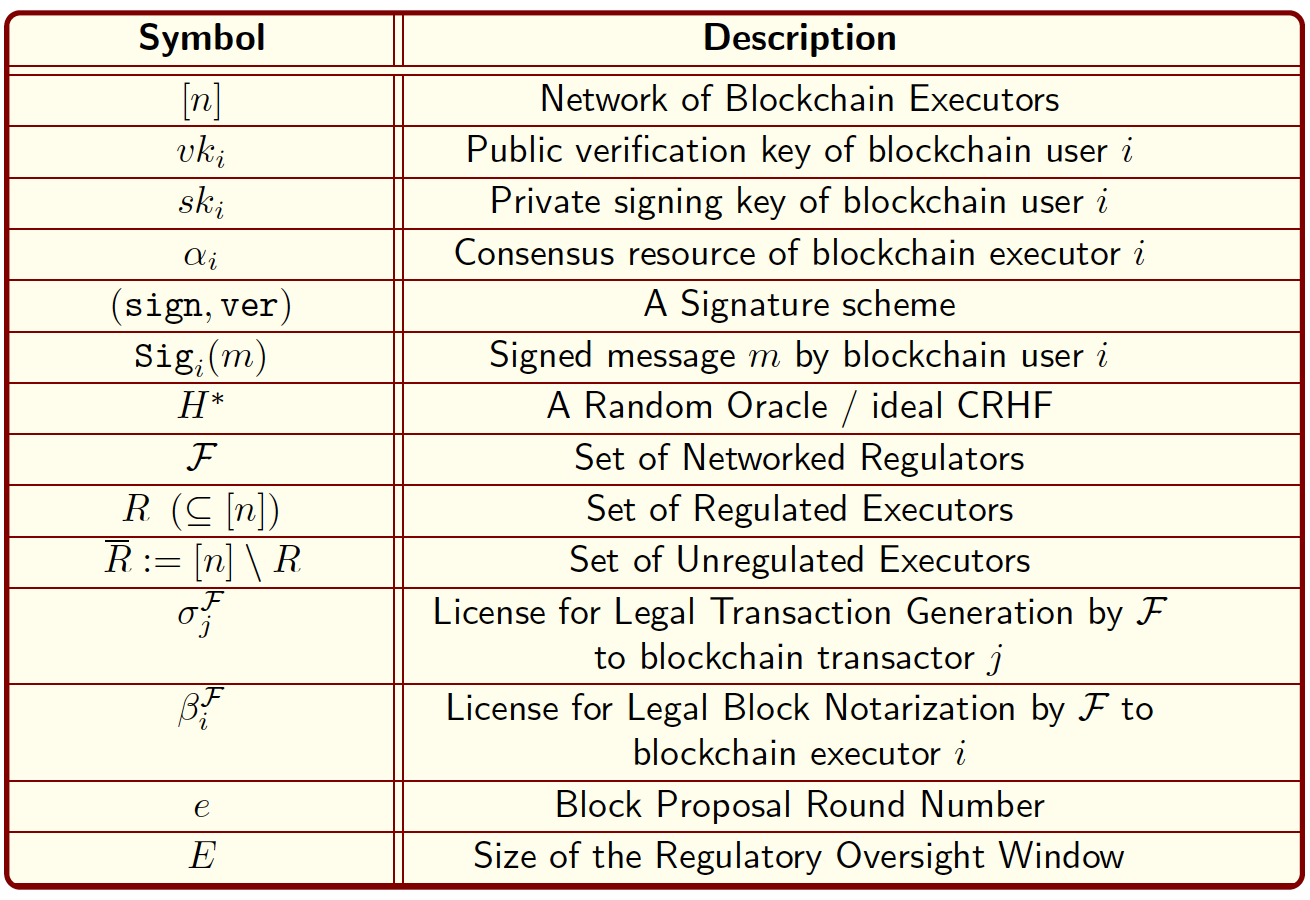}
	\caption{Notation for our Regulatory Framework}
	\vspace{-1em}
	\label{fig:notation}
\end{figure}

\noindent We present the notation that will be used throughout the paper. Given $n$ blockchain consensus protocol executors, we will denote them using $[n] := \{ 1,2, ..., n \}$. The consensus resource of each executor $i \in [n]$ will be given by $\alpha_i$. Given any blockchain user $i$ (may it be a transactor or executor), its private signing key will be denoted by $sk_i$ and its public verification key will be denoted by $vk_i$. We will assume the existence of a signature scheme $(\texttt{sign},\texttt{ver})$, where, for any message $m$, given $\texttt{Sig}_i(m) := (m, \texttt{sign}_{sk_i}(m))$, it is true that $\forall i, \texttt{ver}_{vk_i}(\texttt{Sig}_i(m)) = 1$ (verification succeeds for a signed message). We will also assume the existence of a random oracle $H^*$, realizable through an ideal collision resistant hash function. We will denote the set of networked, cross-jurisdictional regulators by $\mathcal{F}$. Each executor in $[n]$ will be under the jurisdiction of some regulator in $\mathcal{F}$. We will denote the set of regulated executors by $R \hspace*{5pt} (\subseteq [n])$. We will denote the set of unregulated executors with $\mybar{R} \hspace*{5pt} := [n] \setminus R$. We will denote the regulatory licenses for legal transaction proposal by $\sigma^\mathcal{F}_j$ (where $j$ is a transactor) and legal block notarization by $\beta^\mathcal{F}_i$ (where $i$ is an executor). We will denote the block proposal rounds/epochs by $e$. Finally, we will denote the regulatory oversight window, as the number of sequential block proposal rounds for which the regulatory licenses are valid\footnote{Many regulated assets exist whose monetary value and trading terms vary over time. Consequentially, licenses corresponding to these assets may have a date of expiry. The regulator determines this expiry window.}, by $E$. Our notation is summarized in Figure \ref{fig:notation}.

\subsection{Regulated Blockchain System Design Goals}
\label{subsec:rbc-design-goals}

\noindent \emph{1. Enforcing legal policies on blockchain users while preserving anonymity.} \\
\textbf{A1}. As a general rule, the regulators need not intervene in the blockchain network beyond their policymaking role, as is their purview in existing financial institutions \cite{capmarket-reg} and computational systems \cite{nn-regstate}. \\
\textbf{A2}. The identity of the regulated users in the blockchain network should not be unanonymized beyond their legal status in the network: this means that the regulated users preserve their anonymous digital identity, but their digital identity is mapped to a policy statement dictating what are the legal rules applicable to the said identity when employed for cryptocurrency based trade of regulated assets. \\
\textbf{A3}. Regulated users should not be able to participate in the regulated version of the consensus protocol prior to a sanction of their license by the regulatory body. \\

\noindent \emph{2. Requirement of regulatory policy enforcement in lock-step with transaction agreement (blockchain consensus).} \\
As a pre-emptive measure to eliminate illegal transactions within the crypto-economy, regulatory policy enforcement should be in lock-step with the execution of the consensus protocol by regulated executors, so that legal transactions are preferentially agreed over and above the dubious/illegal transactions, and illegal activities through cryptocurrencies are minimized in this way, if not eliminated. This requires re-engineering of existing blockchain protocols, and leads to the following goals: \\
\textbf{B1}. The re-engineered protocol should be different from the original protocol to enable the network to distinguish protocol execution and blocks coming from regulated users following the re-engineered protocol, as opposed to protocol execution and blocks from unregulated users following the original protocol. \\
\textbf{B2}. The design philosophy of the re-engineered blockchain protocol should be same as that of the original blockchain protocol, and the two protocols should be statistically equivalent\footnote{If both protocol states are observed as probability distributions, then the two distributions should be statistically indistinguishable \cite{crypto-kl}.}, so that the re-engineered protocol retains the speed, security and decentralization guarantees of the original protocol. \\
\textbf{B3}. Assuming there are always some legal transactions eligible to be added to the blockchain, the blockchain network being permissionless, must have public guarantees of \emph{legal transaction throughput}: number of legal transactions distributively agreed per unit time. The throughput of legal transactions generated by regulated users should supersede the throughput of dubious transactions generated by unregulated users, in instances of race conditions in block proposal. These guarantees would increase confidence in the legality of the associated crypto-economy. \\
% \textbf{B4}. Given that permissionless blockchain protocols can never achieve full-decentralization without a trusted third party \cite{imposs-fulldecen}, which is discriminatory towards users with low consensus resource power, the regulators can make policies and issue licenses which result in a fully-decentralized, and thus non-discriminatory crypto-economy.

\subsection{Regulated Blockchain Protocol Features}
\label{subsec:rbc-features}
\noindent Given an existing blockchain consensus protocol \textsf{BChain}, we re-engineer the same to define a regulated blockchain protocol \textsf{RBChain}. Under \textsf{RBChain}, the regulatory body $\mathcal{F}$ only licenses (consistent with \textbf{A1,A2,A3}) the consensus protocol transactors and executors to ensure that these blockchain users undertake the responsibility of distributed consensus on legal transactions. Also, the \textsf{RBChain} protocol (defined in Section \ref{sec:reg-proto}) must have the following features:
\begin{itemize}
    \item \emph{Transactions under \textsf{RBChain} should be distinguishable from those under \textsf{BChain}.} This feature would allow the blockchain network to distinguish between legal and dubious transactions by regulated and unregulated transactors respectively.
    \item \emph{Block structure under \textsf{RBChain} should be different from that under \textsf{BChain}.} This feature would allow the blockchain network to distinguish between regulated blocks and unregulated (which could be either of legal or dubious) blocks proposed by regulated and unregulated executors respectively (consistent with \textbf{B1}).
    \item \emph{The execution of \textsf{RBChain} and \textsf{BChain}, when viewed as probability distributions, should be statistically indistinguishable.} This would guarantee that the regulated executors do not have an unfair advantage over unregulated executors, in speed and security during protocol execution (consistent with \textbf{B2}).
    \item \emph{The consensus resource of executors under \textsf{RBChain} should be same as that when they were executing \textsf{BChain}.} This feature would allow the regulated executors to have the same consensus as when they were unregulated, in order to preserve the decentralization status of the blockchain network (consistent with \textbf{B2}).
\end{itemize}

\noindent We address system design goal \textbf{B3} in Section \ref{sec:competition}.

\subsection{The Threat Model in a Regulated Setting}
\noindent We discuss briefly the threat model inherited by \textsf{RBChain}, on being re-engineered from \textsf{BChain}. 
\subsubsection*{Security under the Traditional Byzantine Adversary}
An artefact of the statistical properties of \textsf{RBChain} and \textsf{BChain} being equivalent, we can conclude that \textsf{RBChain} will be secure under a Byzantine fault adversary, under the same security assumptions as applicable to \textsf{BChain} (for example, majority of miners being honest in Bitcoin \cite{bitcoin}, or fraction of money held by honest users being $> \frac{2}{3}$rd of the total wealth in Algorand \cite{algorand}).
\subsubsection*{Block Proposal Competition between Regulated and Unregulated Executors}
Given the permissionless nature of cryptocurrency blockchain networks, these networks can never be wholly regulated. This would entail a strategic competition between legal and dubious block proposal. \\
\noindent In Section \ref{sec:competition}, we study legal transaction throughput as a function of the cumulative consensus resource licensed by the regulatory body $\mathcal{F}$. We derive conditions under which unregulated executors `defect' to a legal block proposal, (i) to avoid legal scrutiny and prosecution; and more importantly, (ii) to have monetary benefit in consensus protocol participation: incentivized to participate in legal block proposal, given dubious block proposal does not give the best reward. We will also assume that regulated executors can form notarization pools and keep notarized blocks private, whereas unregulated executors cannot form such pools due to mistrust among them towards a fair distribution of block notarization reward (consistent with the study of prevalent mistrust among unregulated miners in \cite{btc-minepractice}).

% \draw{Figure: Regulated blockchain system toy model.} \\
\begin{figure}
	\centering
	\includegraphics[width=0.95\linewidth]{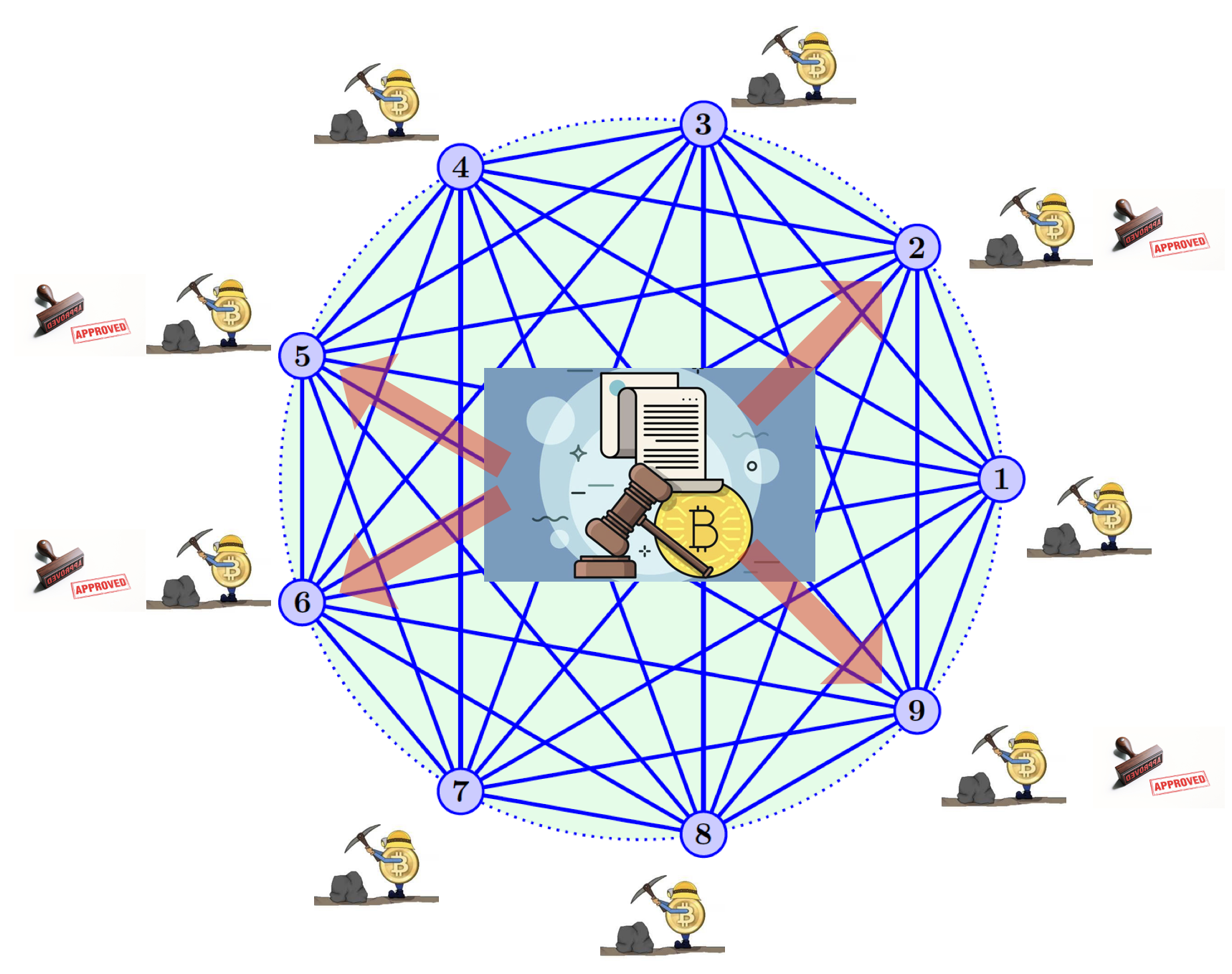}
	\caption{Toy Model of a Permissionless Regulated Blockchain Network. Regulatory body (center) licenses four executors (highlighted with a stamp on the associated coin miner) to propose regulated blocks. The remaining executors are free to propose any type of blocks.}
	\vspace{-1em}
	\label{fig:reg-toy}
\end{figure}

\noindent A toy model of our regulated blockchain system is given in Figure \ref{fig:reg-toy}.

%% file: 04_reg-proto.tex
\section{Regulated Consensus Protocols}
\label{sec:reg-proto}
\noindent We now give exact construction of the regulated version of two popular regimes of blockchain consensus protocols. We will use $\circ$ as a bit-string concatenation operator. To denote statistical indistinguishability of two probability distributions, we will use the notation $\approx_s$.

\subsection{Regulatory Licenses}
Prior to generating verifiably legal transactions and notarizing regulated blocks, regulated blockchain users\footnote{We use the term `users' to refer to both transactors and executors collectively.} wait for license from the regulatory body $\mathcal{F}$ to define the terms for legal trade and regulated block proposal in the blockchain network. \\
\noindent The regulator first announces the legal rules applicable to the cryptocurrency based trade of asset classes $\mathcal{A}$ in regulatory jurisdictions $\mathcal{F}$, through a signed document $\Gamma^\mathcal{F}_\mathcal{A}$. More specifically, $\Gamma^\mathcal{F}_\mathcal{A}$ is a $|\mathcal{F}| \times |\mathcal{A}|$ matrix where each entry $(f,a) \in \mathcal{F} \times \mathcal{A}$ is a list of legal rules pertaining to cryptocurrency based trade of asset $a$ in jurisdiction $f$. These rules may pertain to the legality of asset $a$ in jurisdiction $f$, or the (time varying) valuation of asset $a$ in jurisdiction $f$. Next, the regulator licenses each authorized transactor $j$ by announcing signed permissions for the transactor to trade in jurisdictions $F_j (\subseteq \mathcal{F})$ with asset classes $A_j (\subseteq \mathcal{A})$. This essentially means that under license (denoted by $\sigma^{\mathcal{F}}_j$), transactor $j$ is only permitted to trade under rules $\Gamma^{F^j}_{A_j}$. Finally, given that the regulator approves executors $R \hspace*{3pt} (\subseteq [n])$, the regulator licenses each authorized executor $i \in R$ by announcing signed permissions (denoted by $\beta^{\mathcal{F}}_i$) that $i$ is to include only transactions consistent with $\Gamma^\mathcal{F}_\mathcal{A}$ in the blocks that it proposes. All the licenses announced by the regulator are only valid for an oversight window of $E$ contiguous blocks, starting from some root block $B^{e_0}$ proposed in epoch $e_0$. \\
\noindent The announcement of regulatory licenses is highlighted in Algorithm \ref{alg:reg-licenses}. Note that the legal rules $\Gamma^\mathcal{F}_\mathcal{A}$, and the licenses $\sigma^{\mathcal{F}}_j$ for transactors $j$ and $\beta^{\mathcal{F}}_i$ for executors $i \in [n]$ are only relevant for the regulated protocol specification. They have no bearing on the block proposal competition in the next section, as the said competition only depends on the consensus resource $(\alpha_R,\alpha_{\mybar{R}})$, and the oversight window $E$ (please see Section \ref{sec:competition} for details).

\begin{algorithm}[h]
\caption{Regulatory Licenses}
\label{alg:reg-licenses}
\begin{algorithmic}

\Procedure{$\texttt{RegLicenses}$}{\textsf{BChain}} \\    
\textbf{L0.} Regulator announces over $[n]$: $\texttt{Sig}_{\mathcal{F}}(\textsf{RBChain}, B^{e_0}, E, \Gamma^\mathcal{F}_\mathcal{A})$. \\
\textbf{L1.} For each transactor $j$, regulator issues $\sigma^{\mathcal{F}}_j := \texttt{Sig}_{\mathcal{F}}(\textsf{RBChain}, B^{e_0}, E, vk_j, \textsc{transactor}, F_j, A_j)$. \\
\textbf{L2.} For each executor $i \in R$, regulator issues $\beta^{\mathcal{F}}_i := \texttt{Sig}_{\mathcal{F}}(\textsf{RBChain}, B^{e_0}, E, vk_i, \textsc{executor}, H^*(\Gamma^\mathcal{F}_\mathcal{A}))$.
\EndProcedure

\end{algorithmic}
\end{algorithm}

\subsection{Regulating Proof-of-Work Consensus}

\noindent \emph{Legal Transaction Structure:} Consider the off-chain transfer of a regulated asset of type $a \hspace*{4pt} (\in A_j)$, in jurisdiction $f \hspace*{4pt} (\in F_j)$, to transactor $j$ with a corresponding receipt $\delta^{f,a}_j$. A regulated Bitcoin transaction for transactor $j$ includes $(\sigma^{\mathcal{F}}_j \circ \delta^{f,a}_j)$ in the transaction script. \\
\noindent \emph{The Regulated Bitcoin Protocol:} The Bitcoin \cite{bitcoin} consensus protocol requires the executors (called miners) to solve a compute intensive crypto-puzzle, to notarize a block. Our proposed regulated Bitcoin protocol \textsf{RBitcoin} is given in Algorithm 2, where the regulated miner adds evidence of its license in the coinbase transaction script before trying to solve the crypto-puzzle associated with the regulated block that it wishes to go on-chain.

\begin{algorithm}[h]
\caption{Regulated Bitcoin Consensus Protocol}
\label{alg:reg-btc}
\begin{algorithmic}

\Given{The \textsf{Bitcoin} Consensus Protocol \cite{bitcoin}.
}

\Procedure{\textsf{RBitcoin}}{} \\    
Given a legal block $\mathcal{LB}$, regulated miner $i \in R$ generates a regulated block $\mathcal{RB}$ by adding $H^*(\beta^{\mathcal{F}}_i)$ to the coinbase transaction of $\mathcal{LB}$, and finds nonce $\eta$ such that $H^*(\eta \circ \mathcal{RB})$ lies in the target window.
\EndProcedure

\end{algorithmic}
\end{algorithm}

% \aditya{Argue statistical indistinguishability of Algorithm \ref{alg:reg-btc} vis-a-vis \textsf{Bitcoin}.} \\
\noindent \emph{Security of \textsf{RBitcoin} (Consistent with \textbf{B2} in Section \ref{subsec:rbc-design-goals}):} It is known that the image distribution of the random oracle (ideal CRHF) $H^*$ is uniform, for any pre-image distribution. So, for any block $\mathcal{B}$ mined on for an appropriate nonce $\eta'$ in the traditional \textsf{Bitcoin} consensus protocol, it is true that: $H^*(\eta' \circ \mathcal{B}) \approx_s H^*(\eta \circ \mathcal{RB})$. This implies that mining under \textsf{RBitcoin} is statistically equivalent to mining under \textsf{Bitcoin}.

\subsection{Regulating \textsf{Nxt} Proof-of-Stake}
The \textsf{Nxt} Proof-of-Stake \cite{v-pos} consensus protocol uses the \textsf{IsEligible} deterministic algorithm to elect a validator for notarizing a block. We propose the regulated version of the eligibility algorithm, to validate a regulated block, called \textsf{RIsEligible}, using the regulatory license in the hash pre-image for eligibility proof generation. \textsf{RIsEligible} is given in Algorithm 3. The proof of security of \textsf{RIsEligible} is identical to that of \textsf{RBitcoin}.

\begin{algorithm}[h]
\caption{Regulated \textsf{Nxt} Proof-of-Stake Protocol}
\label{alg:reg-nxtpos}
\begin{algorithmic}

\Given{The \textsf{Nxt} Consensus Protocol \cite{v-pos}.
}

\Procedure{\textsf{NxtPoS-RIsEligible}}{} \\    
Given a legal block $\mathcal{LB}$, regulated validator $i \in R$ receives a regulated block $\mathcal{RB}$ (formed by adding $\beta^{\mathcal{F}}_i$ to $\mathcal{LB}$). $i$ then finds nonce $\eta$ such that $H^*(vk_i \circ \beta^{\mathcal{F}}_i \circ \eta)$ lies in the target window which is a function of time and $\alpha_i$, to become eligible to validate $\mathcal{RB}$.
\EndProcedure

\end{algorithmic}
\end{algorithm}

% ***DELETED***
\begin{comment}
\subsection{Regulating the BA* Protocol}

\begin{algorithm}[h]
% \renewcommand{\thealgorithm}{}
\caption{Regulated BA$^*$ Protocol}
\label{alg:reg-algo}
\begin{algorithmic}

\Given{The \textsf{Algorand} Consensus Protocol \cite{algorand}.}

\Procedure{\textsf{RAlgorand}}{} \\    
Given a legal block $\mathcal{RB}$ proposed by a regulated minter $i \in R$, each honest minter $j \in R^*$ with an \textsf{Algorand} signing key $sk_j$, votes on $H^*(\mathcal{RB})$ employing $VRF_{sk_j}(seed \circ role \circ \beta^{\mathcal{F}}_i)$, for every $seed$ and $role$.
\EndProcedure

\end{algorithmic}
\end{algorithm}

% \aditya{Argue statistical indistinguishability of Algorithm \ref{alg:reg-algo} vis-a-vis \textsf{Algorand}.} \\
\noindent \emph{Security of \textsf{RAlgorand} (Consistent with \textbf{B2} in Section \ref{subsec:rbc-design-goals}):} For the VRFs used in \textsf{Algorand}, it is true that for any two bit strings $x, x'$, and any given signing key $sk$, $VRF_{sk}(x) \approx_s VRF_{sk}(x')$ \cite{algorand}. Consequently, it is true that $VRF_{sk_j}(seed \circ role) \approx_s VRF_{sk_j}(seed \circ role \circ \beta^{\mathcal{F}}_i)$, and \textsf{Algorand} and \textsf{RAlgorand} have statistically indistinguishable block validation votes' distribution.
\end{comment}

%% file: 05_comp-analysis.tex
\section{Block Proposal Competition between Regulated and Unregulated Executors}
\label{sec:competition}

% \aditya{Competition modelled as a two-player blockchain stochastic game, inspired from \cite{bmg,bmg-pf}.}
We now analyze the competition between regulated and unregulated consensus protocol executors (as per goal \textbf{B3} in Section \ref{subsec:rbc-design-goals}) as a two player stochastic game, inspired from \cite{bmg,bmg-pf}. We formally define the stochastic game, and state best responses by the regulated executors $R$ and unregulated executors $\mybar{R}$ (in the form of their Nash Equilibria \cite{sg-solan,bmg} strategies), as a function of the total consensus resource regulated by $\mathcal{F}$. \\

\noindent The proof of each theorem in this section, is given in Appendix \ref{app:thm-proofs}.

\subsection{The Regulated Blockchain Game Features}

\noindent We give the general characteristics and an overview of our regulated blockchain stochastic games, in this subsection.

\subsubsection*{Generality of Blockchain Mining Games}
The blockchain mining games introduced by Kiayias and Koutsoupias \cite{bmg,bmg-pf}, atop which our regulated blockchain stochastic games are constructed, allow a minority resource consensus executor $\mathbf{p} \hspace*{3pt}$ ($\subseteq [n]$, with $\alpha_{\mathbf{p}} < 0.5$) to consider switching to an interior block (a block that is not the most recent block), or the frontier block, in the competing chain, as a better response than its present notarization strategy. However, the pioneering selfish mining strategy SM1 \cite{btc-sm1} (named in \cite{btc-osm}) only allows $\mathbf{p}$ to switch to the frontier block in the competing chain. Consequently, the mining games introduced in \cite{bmg,bmg-pf} are more general in their strategy than the original selfish mining proposal.

\subsubsection*{Preliminaries}
\begin{figure}
	\centering
	\includegraphics[width=1.0\linewidth]{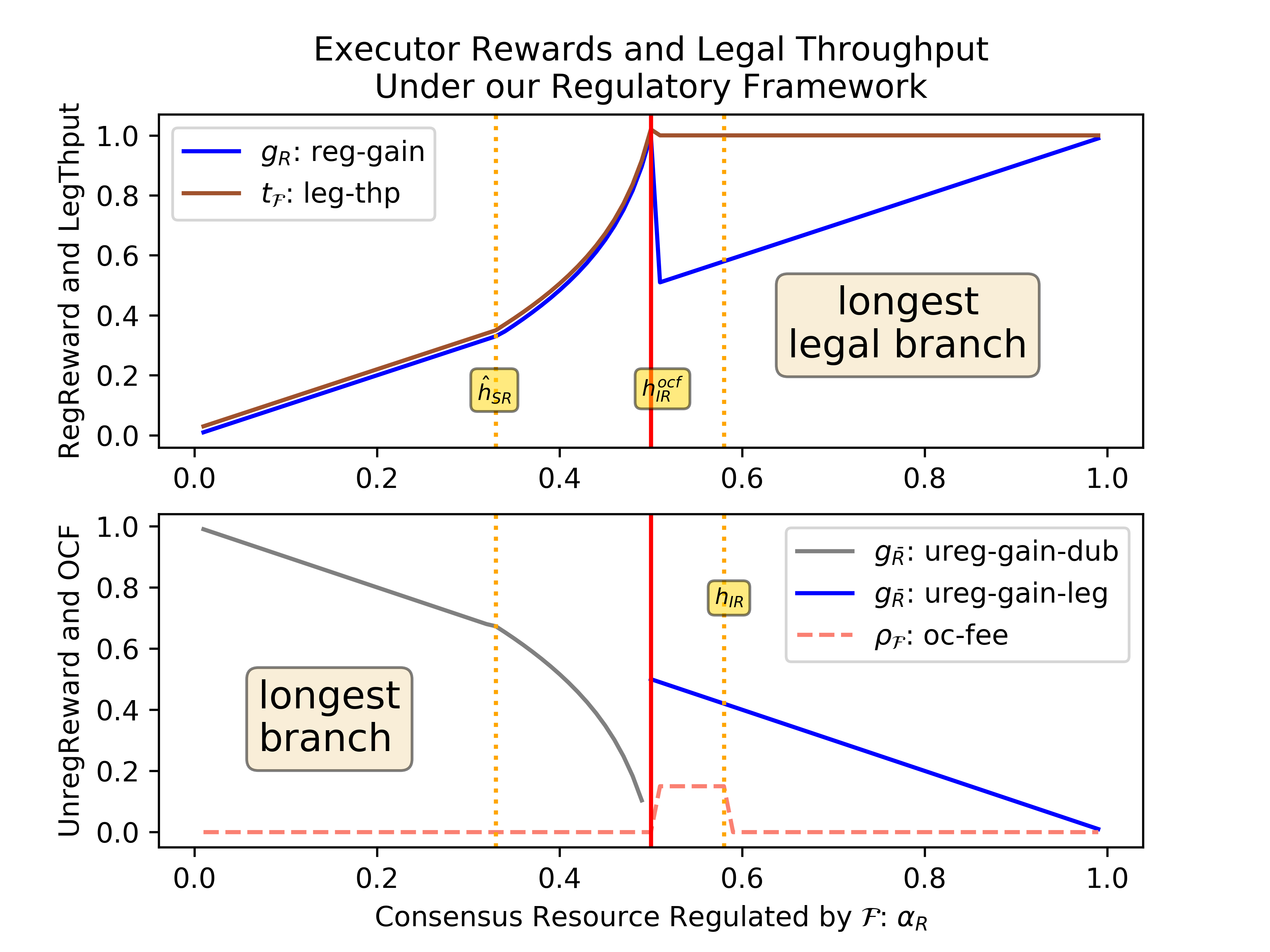}
	\caption{A summary of the results of our regulatory system. When $\alpha_R \leq 0.5$, the legal transaction throughput is sub-optimal under the longest branch rule. However, when $\alpha_R > 0.5$, the longest legal branch rule wins, unregulated executors can do no better than proposing legal blocks, and the legal transaction throughput is maximized. Consensus thresholds $h_{IR}, h^{ocf}_{IR}$, and $\hat{h}_{SR}$ are formally defined later.}
	\vspace{-1em}
	\label{fig:rbc-results}
\end{figure}

\noindent Our blockchain notarization stochastic game defines competition between two player categories: the regulated executors $R$ and the unregulated executors $\mybar{R}$. We assume that the total consensus resource in the blockchain network is normalized: $\sum_{i \in [n]} \alpha_i = 1$; and the expected revenue / reward per epoch of the blockchain for any category of players $\mathbf{p} \in \{ R, \mybar{R} \}$ is denoted by $g_{\mathbf{p}}$, and is a function of $\alpha_{\mathbf{p}} = \sum_{i \in \mathbf{p}} \alpha_i$. Unless the regulator $\mathcal{F}$ adds extra transactions to the blockchain (Section \ref{subsec:sg-ir-ocf}), $g_R + g_{\mybar{R}} = 1$. Our game evolves as a block tree of width two, with one branch consisting of regulated blocks notarized by $R$, and the other branch consisting of legal or dubious blocks notarized by $\mybar{R}$, and the forking point of the block tree has a root block, say $B^{e_0}$. If either of $R$ or $\mybar{R}$ abandons its branch for notarizing on a block linked to its competing branch, the root block $B^{e_0}$ moves accordingly to newly selected block for notarization, and the game starts afresh. We also allow the regulator $\mathcal{F}$ to add an additional reward $\rho_{\mathcal{F}}$ in the regulated blocks, when $R$ is in a small majority in the blockchain network ($0.5 < \alpha_R < 0.58$), resulting in $g_R + g_{\mybar{R}} \geq 1$. Our game has depth $E$, equal to the oversight window and the window for collecting the notarization reward (coinbase reward in Bitcoin \cite{bmg-pf}). The game depth $E$ means that the first branch originating from $B^{e_0}$ that achieves height $E$ is confirmed as part of the blockchain, the other competing branch is orphaned, and again the game starts afresh. Finally, we assume that the executors collect their block notarization reward at the end of the game depth (for example, for $E = 100$ in Bitcoin \cite{bmg}). For a simplified analysis of the game, we assume $E = \infty$, unless stated otherwise. \\
\noindent We will denote the normalized legal transaction throughput, which can also be interpreted as the expected number of legal blocks agreed at each epoch of the blockchain, by $t_{\mathcal{F}}$. Given that regulated executors only propose legal transactions, and unregulated executors may propose either of legal or dubious transactions, it is easy to see that $t_{\mathcal{F}} \geq g_R$. \\
\noindent \emph{Stochastic Game Summary}. Our results on normalized legal transaction throughput $t_{\mathcal{F}}$ and normalized expected block rewards for regulated executors $g_R$ and unregulated executors $g_{\mybar{R}}$ are summarized in Figure \ref{fig:rbc-results}. These results are formally detailed in Section \ref{subsec:sg-ir}, Section \ref{subsec:sg-ir-ocf}, and Section \ref{subsec:sg-sr}.

% \draw{Figure: Competition Model Results.} \\

\noindent We now define \emph{how} executors choose to release their notarized blocks, and which blocks they may select to notarize future blocks on.

\subsubsection*{Block Release Models}

\noindent We first define the block release models (from \cite{bmg,bmg-pf}), which elucidate \emph{when} consensus executors choose to reveal information about their notarized blocks. \\

\noindent The first block release model stated below can be adopted by both regulated executors $R$ and unregulated executors $\mybar{R}$.
\begin{definition}[Immediate Release Model]
A consensus executor follows the \emph{immediate block release (IR) model} when, any block notarized by it is immediately released and added to the blockchain for use by other executors. \\
\end{definition}

\noindent Notarization rewards earned by a notarization pool must be distributed among pool members appropriately. Due to prevalent distrust between unregulated executors $\mybar{R}$ in the fair distribution of notarization rewards among members of a secret notarization pool created by them \cite{btc-minepractice}, the second block release model stated below can be adopted by regulated executors $R$ alone. Here, the administrator denotes the pool leader.
\begin{definition}[Strategic Release Model]
A consensus executor follows the \emph{strategic block release (SR) model} when, on successful notarization of block(s) by it, the (block administering) executor \emph{announces} its existence, but the block(s) can only be used by other executors when the administrator decides to \emph{release} them. 
\end{definition}

\subsubsection*{Consensus Execution Strategies}

\noindent The honest strategy where a consensus executor notarizes blocks at the end of the longest existing unconfirmed branch of the blockchain, is traditionally referred to as the \texttt{Frontier} strategy \cite{bmg,bmg-pf}. Under the \texttt{Frontier} strategy, the expected gain per epoch of the associated executor is equal to the consensus resource possessed by it. We now give definitions of equivalent (to \texttt{Frontier}) consensus execution strategies in our regulated setting. We will use $\mathcal{DB}$ to denote dubious blocks, $\mathcal{LB}$ to denote legal blocks, and $\mathcal{RB}$ to denote regulated blocks. \\

\noindent We first define the frontier block notarization strategies that the unregulated executors $\mybar{R}$ can adopt.

\begin{definition}[\texttt{DubFrontier}]
An unregulated consensus executor follows the \texttt{DubFrontier} strategy, when it notarizes legal ($\mathcal{LB}$) or dubious ($\mathcal{DB}$) blocks chained at the frontier block of the longest dubious branch of the blockchain.
\end{definition}

\begin{definition}[\texttt{LegFrontier}]
An unregulated consensus executor follows the \texttt{LegFrontier} strategy, when it notarizes legal ($\mathcal{LB}$) blocks chained at the frontier block of the longest legal branch of the blockchain. \\
\end{definition}

\noindent We now define the frontier block notarization strategies that the regulated executors $R$ can adopt. We will also assume that the regulatory body $\mathcal{F}$ may adopt a pay-forward scheme (explained in Section \ref{subsec:sg-ir-ocf}) in the regulated blocks.

\begin{definition}[\texttt{RegFrontier}]
A regulated consensus executor follows the \texttt{RegFrontier} strategy, when it notarizes regulated ($\mathcal{RB}$) blocks chained at the frontier block of the longest legal branch of the blockchain.
\end{definition}

\begin{definition}[\texttt{RegFrontier}($\rho_\mathcal{F}$)]
A regulated consensus executor follows the \texttt{RegFrontier}($\rho_\mathcal{F}$) strategy, when it notarizes regulated ($\mathcal{RB}$) blocks chained at the frontier block of the longest legal branch of the blockchain, with a pay-forward of $\rho_\mathcal{F}$ (which is a function of $\alpha_R$) by the regulatory body $\mathcal{F}$ in the regulated blocks, as an oversight compliance fee (OCF).
\end{definition}

\begin{definition}[\texttt{RDubFrontier}]
A regulated consensus executor follows the \texttt{RDubFrontier} strategy, when it notarizes regulated ($\mathcal{RB}$) blocks chained at the frontier block of the longest dubious branch of the blockchain. \\
\end{definition}

% \begin{definition}[\texttt{BiRegFrontier}($\rho_\mathcal{F}$)]
% A consensus executor follows the \texttt{BiRegFrontier}($\rho_\mathcal{F}$) strategy, when it notarizes legal blocks chained at the frontier block of the unconfirmed but notarized legal branch of the blockchain, where each block in the legal branch has an OCF of $\rho_\mathcal{F}$ or zero. \\
% \end{definition}

\noindent \emph{The longest legal branch rule.} Block notarization strategies \texttt{LegFrontier}, \texttt{RegFrontier} and \texttt{RegFrontier}($\rho_\mathcal{F}$) constitute notarization on the longest legal branch: at every epoch, all executors following these strategies add legal blocks to the longest existing branch of legal blocks. \\
\noindent \emph{Our competition analysis.} In the following analysis, the first two games (in Section \ref{subsec:sg-ir} and Section \ref{subsec:sg-ir-ocf}), deal with deriving the conditions under which attacks by the unregulated executors on the legal branch fail, given that the regulator licensed the majority of the consensus resource in the blockchain network. The third and final game (in Section \ref{subsec:sg-sr}) deals with deriving conditions under which attacks by the regulated executors on the dubious branch succeed, given that the unregulated consensus resource in the blockchain network is in a majority.

\subsection{A Stochastic Game with Immediate Block Release}
\label{subsec:sg-ir}

\begin{figure}
	% \centering
	\includegraphics[width=1.0\linewidth]{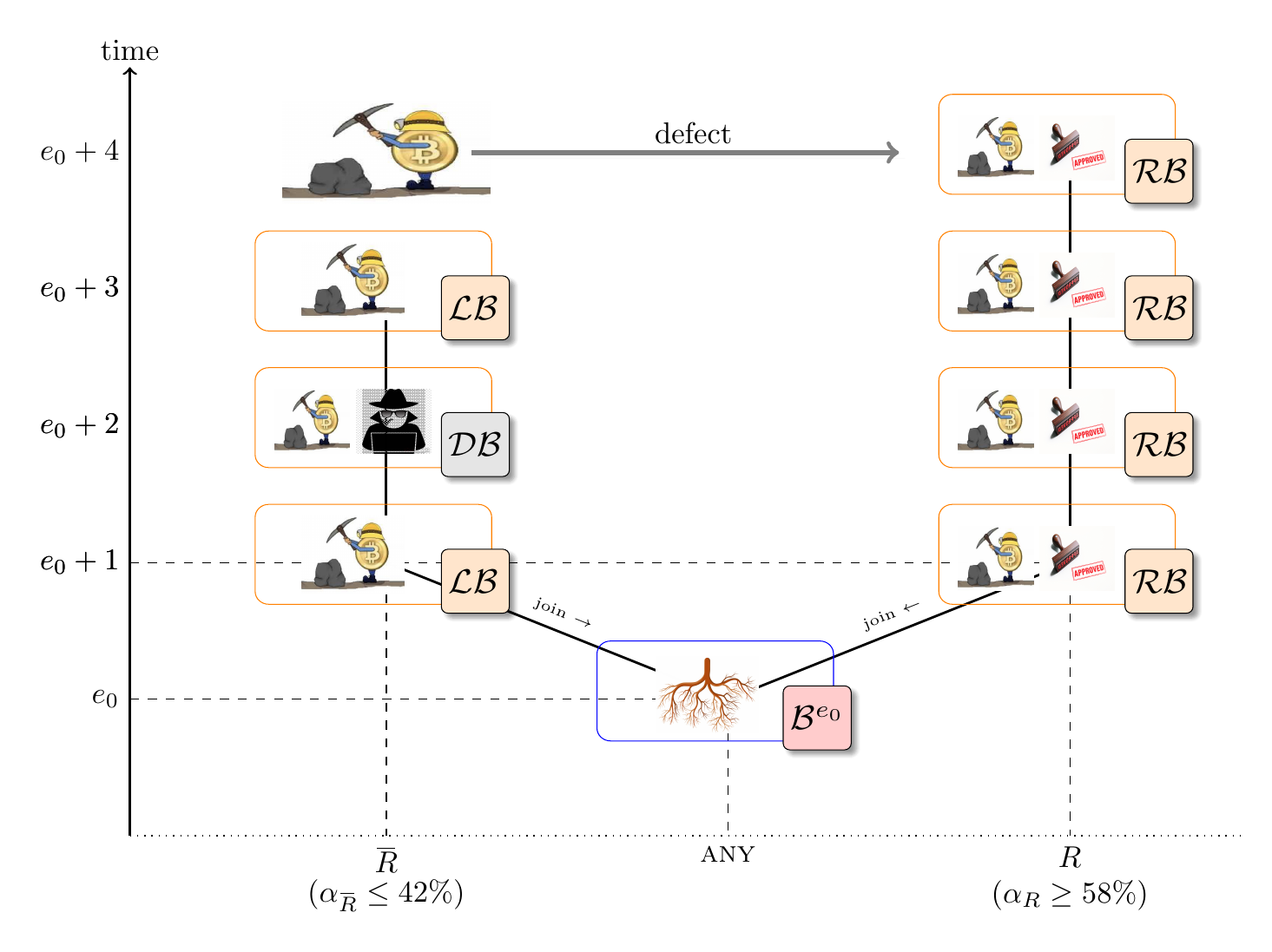}
	\caption{When $\geq 58\%$ of the consensus resource in the blockchain network is regulated, the legal branch (right) corresponding to the regulated executors $R$ wins, by forcing $\mybar{R}$ to abandon and defect from their branch. }
	\vspace{-1em}
	\label{fig:R-free-dom}
\end{figure}

Our first two player game consists of competition between the regulated executors $R$ and the unregulated executors $\mybar{R}$, when both players notarize and release their blocks immediately, the regulator $\mathcal{F}$ licenses more than $1 - h_{IR} = 58\%$ of the consensus resource in the blockchain network, and the regulated executors add regulated blocks to the existing longest legal branch. The best responses (in terms of expected gain maximization) for both players are formalized in the following theorem (depicted in Figure \ref{fig:R-free-dom}). \\

% \draw{Figure: Regulated Executor Dominance.} \\

\begin{theorem}[Equilibrium under IR]
In the immediate block release (IR) model, given regulated executors $R$ follow the `longest legal branch' rule, and $\alpha_{\mybar{R}} < h_{IR} = 0.42$, then $R$ playing \texttt{RegFrontier} and $\mybar{R}$ playing \texttt{LegFrontier} is a Nash Equilibrium.
\end{theorem}
\emph{Theorem Implication:} This theorem implies that, if the regulatory body $\mathcal{F}$ is successful in licensing more than $58\%$ of the total consensus resource, then no executor can do better than adding legal blocks at the frontier block of the legal notarized but unconfirmed branch in the blockchain, resulting in a fair block reward for each executor type: $g_R = \alpha_R, g_{\mybar{R}} = \alpha_{\mybar{R}}$, and a $100\%$ legal transaction throughput in the blockchain network: $t_{\mathcal{F}} = 1$. \\

\begin{figure}
	\centering
	\includegraphics[width=0.95\linewidth]{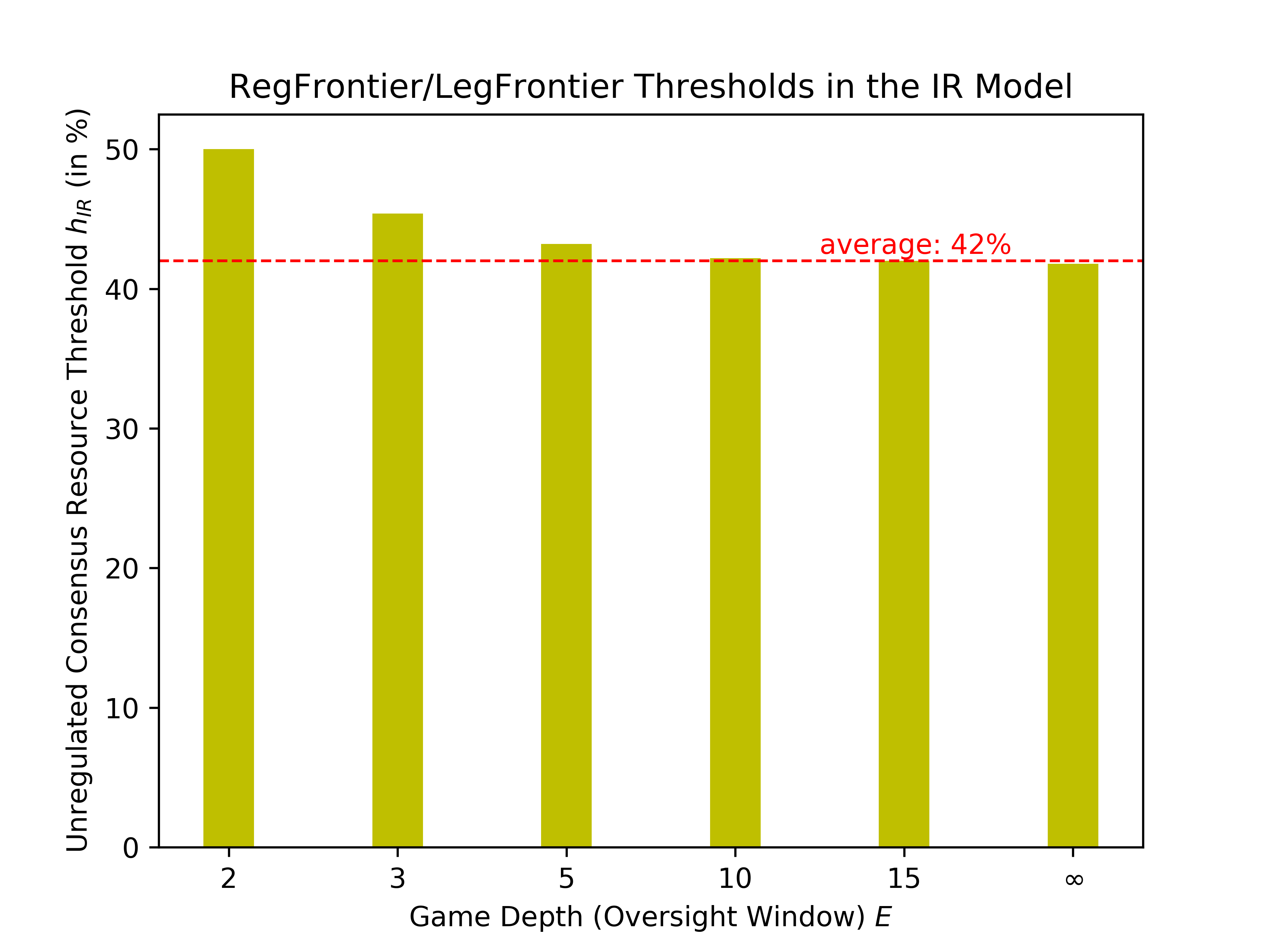}
	\caption{Upperbound on the Unregulated Consensus Resource for the RegFrontier/LegFrontier strategies, as a function of the Game Depth.}
	\vspace{-1em}
	\label{fig:hIR-E}
\end{figure}

\noindent The consensus resource bound $h_{IR}$ on unregulated executors under immediate release, is a function of the game depth (oversight window) $E$, with experimental values given in Figure \ref{fig:hIR-E}, with an approximate value of $0.42$ (please see Section \ref{subsec:sg-thres} for details).  

% \draw{Plot/Table: Unregulated Executor Defection Threshold $h_{IR}$ as a function of the Oversight Window $E$.}

\subsection{A Stochastic Game with Immediate Block Release and an Oversight Compliance Fee}
\label{subsec:sg-ir-ocf}

\begin{figure}
	% \centering
	\includegraphics[width=1.0\linewidth]{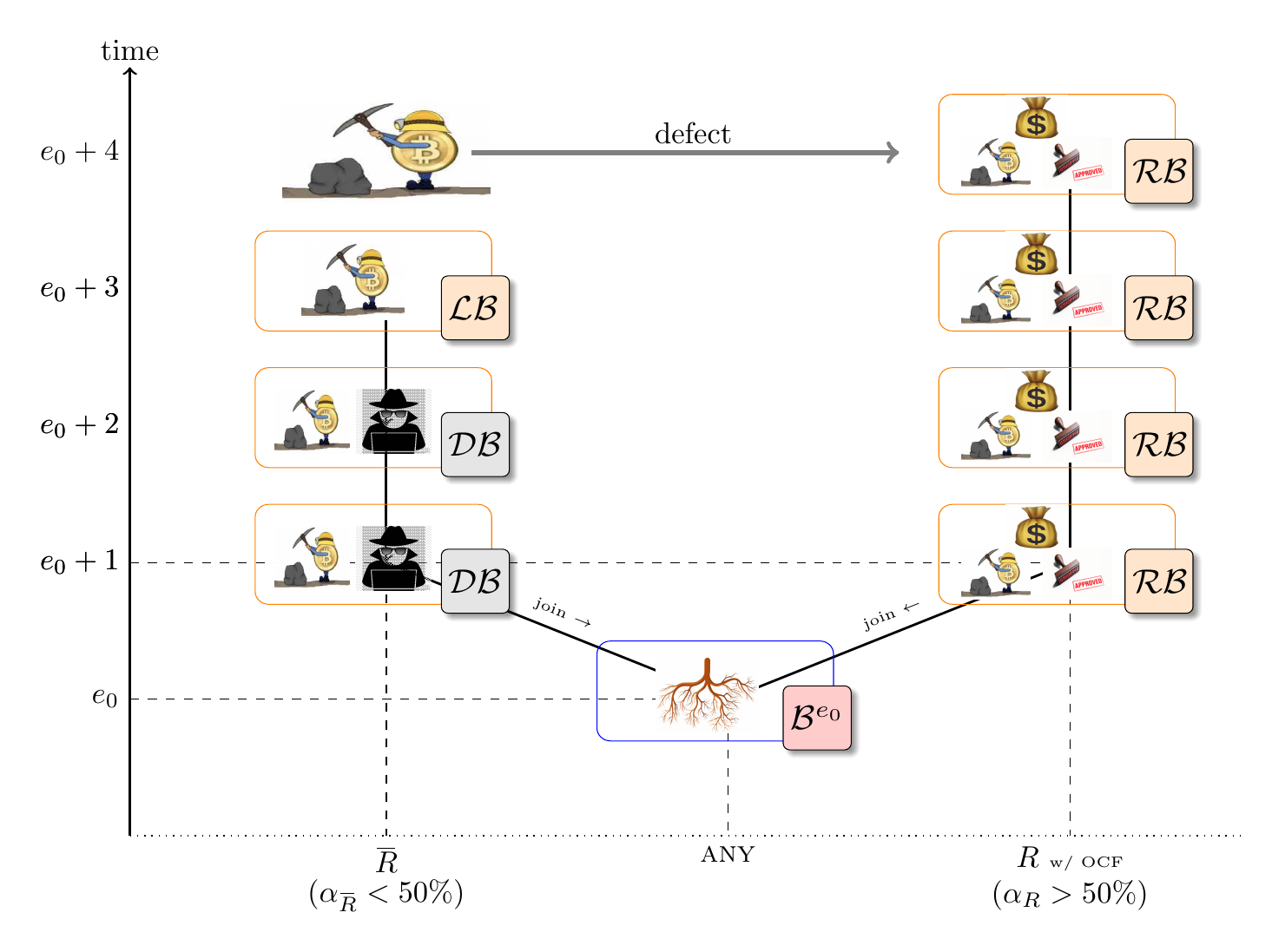}
	\caption{When $> 50\%$ of the consensus resource in the blockchain network is regulated, and the regulator adds an OCF (denoted by the money bag in the $\mathcal{RB}$ blocks), the legal branch (right) corresponding to the regulated executors $R$ wins, by forcing $\mybar{R}$ to abandon and defect from their branch.}
	\vspace{-1em}
	\label{fig:R-paid-dom}
\end{figure}

Our second two player game consists of competition between the regulated executors $R$ and the unregulated executors $\mybar{R}$, when both players notarize and release their blocks immediately, the regulator $\mathcal{F}$ licenses more than $h^{ocf}_{IR} = 0.50$ of the consensus resource in the blockchain network. In this game, the regulator additionally remits an oversight compliance fee (OCF) $\rho_{\mathcal{F}}$ as an extra transaction in each regulated block (to incentivize legal block proposal over dubious block proposal), which is claimed by the executor corresponding to the confirmed block following the said regulated block: given that some block $B^e$ is regulated and contains the OCF, then notarizer of block $B^{e+1}$ claims the said OCF. This OCF is a function of $\alpha_R$. Here again, the best responses for both players, given that the regulated executors add regulated blocks to the existing longest legal branch, are formalized in the following theorem (depicted in Figure \ref{fig:R-paid-dom}). \\

% \draw{Figure: Regulated Executor Dominance under Oversight Compliance Fee.}
\begin{theorem}[Equilibrium under IR with an OCF]
In the immediate block release (IR) model, given regulated executors $R$ follow the `longest legal branch' rule, and $\alpha_R > h^{ocf}_{IR} = 0.50$, then $R$ playing \texttt{RegFrontier($\rho_\mathcal{F}$)} and $\mybar{R}$ playing \texttt{LegFrontier} is a Nash Equilibrium.
\end{theorem}
\emph{Theorem Implication (in conjunction with Theorem 1):} As a consequence of this theorem and the previous one, if the regulatory body $\mathcal{F}$ is successful in licensing between $50\%$ and $58\%$ of the total consensus resource, \emph{and} mandates legal blocks to an oversight compliance transaction fee $\rho_\mathcal{F}$ (which is paid by $\mathcal{F}$ and is a function of $\alpha_R$), then again, no executor can do better than adding legal blocks at the frontier block of the legal branch in the blockchain, resulting in a fair block reward for each executor type: $g_R \geq \alpha_R, g_{\mybar{R}} \geq \alpha_{\mybar{R}}$ (inequality due to $\rho_\mathcal{F}$), and a $100\%$ legal transaction throughput in the blockchain network: $t_{\mathcal{F}} = 1$. Note that when the regulated consensus resource in the blockchain network is between $50\%$ and $58\%$, if the regulator $\mathcal{F}$ does not include the OCF in the regulated blocks, it would still be true that $t_{\mathcal{F}} = 1$, but the unregulated executors $\mybar{R}$ can attack and successfully add legal blocks linked to an interior block in their competing branch, resulting in $g_R < \alpha_R$. Thus it is imperative that the regulator adds the OCF in the regulated blocks, to save the expected gain $g_R \hspace*{5pt} (\geq \alpha_R)$ of the regulated executors $R$. \\

% \begin{theorem}[Equilibrium under IR with an OCF and a short Oversight Window]
% In the immediate block release model, given $\alpha_R \geq 0.56$ and $E = 8$, $R$ playing \texttt{BiRegFrontier($\rho_\mathcal{F}$)} and $\mybar{R}$ playing \texttt{RegFrontier(0)} is a Nash Equilibrium.
% \end{theorem}
% \emph{Theorem Implication (in conjunction with Theorem 1):} If the regulatory body $\mathcal{F}$ licenses between $56\%$ and $58\%$ of the total consensus resource, as a consequence of this theorem and theorem 1, then the unregulated executors can do no better than adding legal blocks with no OCF transactions at the tip of the legal branch, and the regulated executors can do no better than adding legal blocks with a mix of both OCF transactions and no OCF transactions at the tip of the legal branch. This again results in a $100\%$ legal transaction throughput in the blockchain network. \\

\noindent \emph{Justification for a regulatory pay-forward.} We justify this model, as an answer to the following question: given that regulated executors follow the pay-forward scheme with money pitched in by the regulators on the legal branch, why cannot there be a similar pay-forward scheme enforced by unregulated executors on the dubious branch? \\
Here we will assume that world governments, and consequently the regulatory bodies, are richer than the executors and want to preserve the revenue generated by their regulated executors (in order to motivate licensing within the blockchain network). If the unregulated miners even decide to pitch some money per block for the unregulated branch, say $w_{\mybar{R}}$, then the regulators can counterbalance the legal branch by pitching $w_R = w_{\mybar{R}} + \rho_\mathcal{F}$, where $\rho_\mathcal{F}$ is defined as a function of $\alpha_R$ before. This way, the branch of the regulated executors will have an additional weight of $\rho_\mathcal{F}$, and our model is equivalent to the blockchain pay-forward mining game in \cite{bmg-pf}. Consequently the results of that game apply.

\subsection{A Stochastic Game with Strategic Block Release by Regulated Executors}
\label{subsec:sg-sr}

Our final stochastic game considers best responses by $R$ and $\mybar{R}$, given that the regulator $\mathcal{F}$ licenses less than one-third ($\hat{h}_{SR}$ = 0.33) of the total consensus resource, the unregulated executors (being in a majority) follow the longest branch rule, and the regulated executors can follow the strategic block release (SR) model. The best responses are given in the following theorem. \\

\begin{figure}
	% \centering
	\includegraphics[width=1.0\linewidth]{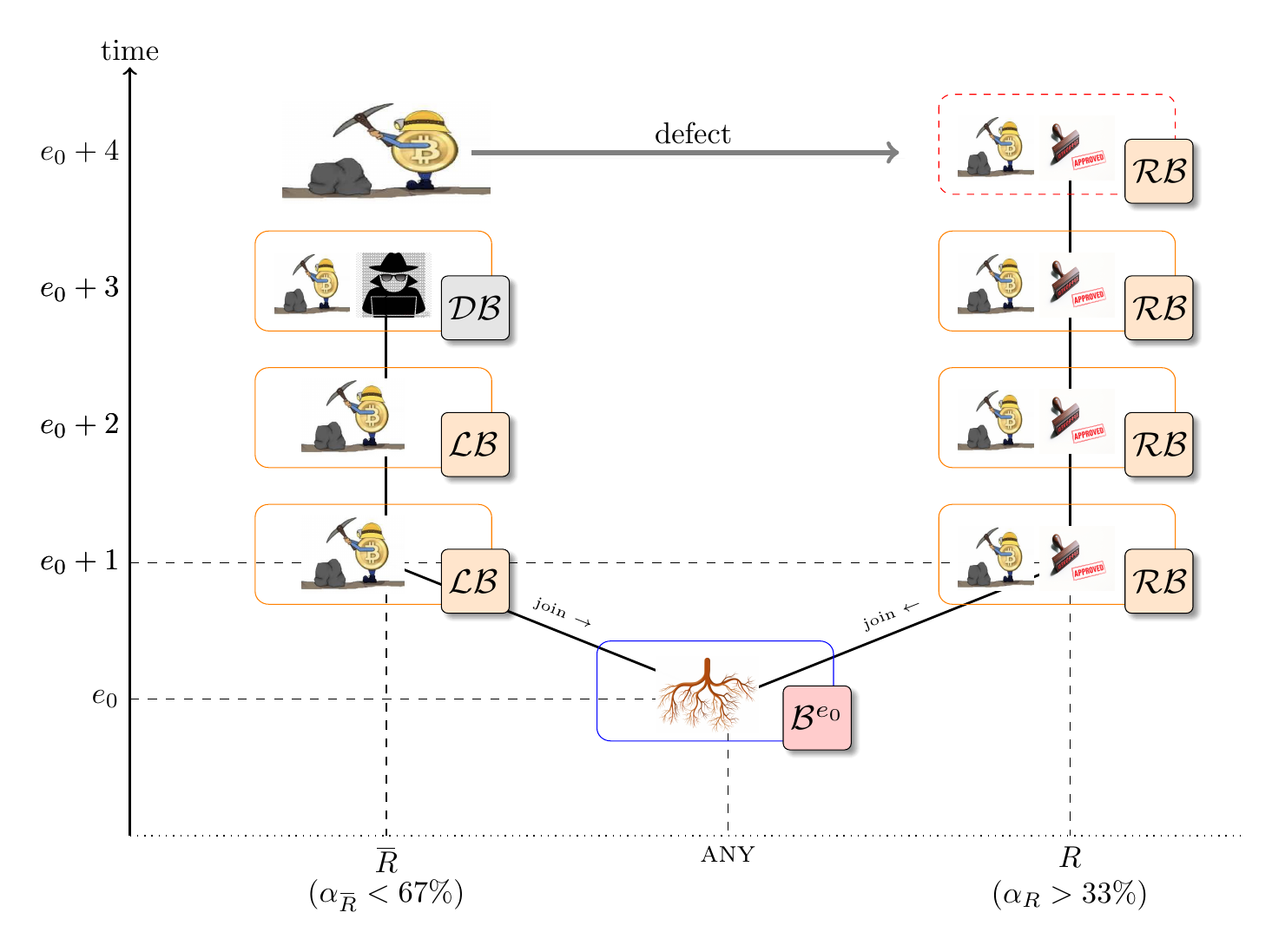}
	\caption{When $> 33\%$ but less than a majority of the consensus resource in the blockchain network is regulated, the regulated executors $R$ can force the unregulated executors $\mybar{R}$ to abandon and defect from their branch through strategic release (unreleased $\mathcal{RB}$ block denoted by a dashed box in epoch $e_0+4$).}
	\vspace{-1em}
	\label{fig:R-prune}
\end{figure}

\begin{theorem}[Equilibrium under SR from $R$]
In the strategic block release (SR) model for regulated executors $R$, given unregulated executors $\mybar{R}$ play \texttt{DubFrontier} with immediate block release (IR), $R$ playing \texttt{RDubFrontier} is a Nash Equilibrium only if $\alpha_R \leq \hat{h}_{SR} = 0.33$.
\end{theorem}
\emph{Theorem Implication (in conjunction with Theorem 2):} The legal transaction throughput in the dubious branch of the blockchain, without loss of generality, would be less than $100\%$. In the case that the regulatory body $\mathcal{F}$ licenses less than $33\%$ of the total consensus resource, the regulated executors can do no better than adding legal blocks at the tip of the dubious branch, inducing a small increase in the legal transaction throughput. However, if the regulatory body $\mathcal{F}$ is successful in licensing more than $33\%$ (and less than $50\%$) of the total consensus resource, it can orphan some of the dubious blocks through strategic release (or equivalently using the original selfish mining attack \cite{btc-sm1}), thereby inducing a higher increase in the legal transaction throughput: $t_{\mathcal{F}} \geq g_R > \alpha_R$ (and also $g_{\mybar{R}} < \alpha_{\mybar{R}}$). This scenario is depicted in Figure \ref{fig:R-prune}. \\

% \draw{Figure: Regulated Executor Shave-Off Dubious Blocks.}

\noindent \emph{Justification of SR by regulated executors alone.} Given that the regulated executors follow strategic release, we argue next that the unregulated executors cannot form a competing pool of their own and also follow strategic release. \\
Remember that executors on the unregulated branch are of two types: law-abiding (but not licensed by the regulator), and law-breaking. It is clear from the Silk road study (Section \ref{sec:casestudy}), that a major reason why law breakers choose Bitcoin for illegal trade is to misuse the anonymity it provides. If the law breaking executors decide to form an untrustworthy notarization (mining) pool with unlicensed law abiding executors for strategic release, it has been established in a study \cite{btc-minepractice} that such pools will not sustain due to mistrust that a dishonest notarization pool administrator may not fairly distribute notarization rewards to pool members. So the chances of sustainable mining/validation pools forming on the dubious branch among these untrustworthy executors are low.

\subsubsection*{Regulated Executors (Righteously) Attacking the Blockchain}

\noindent In the instance that the regulated executors $R$ are in a majority in the blockchain network, with $\alpha_R > 0.5$, these executors might attack the blockchain by using selfish mining strategies from \cite{btc-sm1,btc-osm,bmg}, thereby ensuring $g_R = 1 > \alpha_R$, and consequently achieving the regulator $\mathcal{F}$'s goal of $t_{\mathcal{F}} = 1$. However, this policy is controversial and socially unacceptable for two reasons: (i) this requires hijacking the blockchain by a majority of the consensus executors, and more importantly (ii) it kills the legal transactions notarized in any competing branch. So it is prudent to employ the `longest legal branch' rule when the regulated executors are in a majority, thereby fairly maintaining $g_R = \alpha_R$, and achieving $t_{\mathcal{F}} = 1$. \\
\noindent In the instance $0.33 < \alpha_R < 0.5$ as discussed in this subsection, assuming that the unregulated executors $\mybar{R}$ might misuse their dominance in the blockchain network to notarize dubious transactions, the regulated executors may resort to a white hat attack on the blockchain by adopting one of the selfish mining strategies from \cite{btc-sm1,btc-osm,bmg} to ensure $t_{\mathcal{F}} \geq g_R > \alpha_R$.

\subsection{The Consensus Resource Thresholds}
\label{subsec:sg-thres}
Our stochastic game for block notarization competition between regulated and unregulated executors is equivalent in analysis to the seminal blockchain mining games proposed in \cite{bmg,bmg-pf}, and consequently the results of those game models directly apply. The consensus resource thresholds $h_{IR}, h^{ocf}_{IR}$, and $\hat{h}_{SR}$ are rigorously established in \cite{bmg,bmg-pf}. In \cite{bmg}, it was proven that $0.361 \leq h_{IR} \leq 0.455$, but experimentally $h_{IR} \approx 0.42$. Similarly \cite{bmg} proves $\hat{h}_{SR} \geq 0.308$, but in \cite{btc-osm}, it is experimentally shown that $\hat{h}_{SR} \approx 0.33$. Finally, it is proven in \cite{bmg-pf}, that $h^{ocf}_{IR} = 0.5$.

\subsection{Tying the Results with Existing Protocols}

\noindent We now interpret our blockchain notarization stochastic game in the context of two prominent consensus protocol regimes: Proof-of-Work from Bitcoin and vanilla Proof-of-Stake.

\subsubsection*{Competition in Bitcoin}
Our regulated blockchain stochastic game is inspired from \cite{bmg,bmg-pf} where the analysis is focused on game-theoretically determining optimal mining strategies for the Bitcoin consensus protocol in an unregulated setting. In the Bitcoin consensus protocol, $\alpha_R$ would represent the total hash power regulated by $\mathcal{F}$ in the Bitcoin network. Although the game depth for Bitcoin is $100$ blocks as determined by the coinbase reward rule of the protocol \cite{bmg}, all the consensus resource thresholds apply, verifiable through results in \cite{bmg,bmg-pf,btc-osm}, by replacing $E = \infty$ with $E = 100$. For example, for an approximate hash power upper-bound $h_{IR}$ on $\alpha_{\mybar{R}}$, extrapolated for $E = 100$ through experimental results in \cite{bmg}, please see Figure \ref{fig:hIR-E}.

\subsubsection*{Competition in vanilla Proof-of-Stake}

\noindent Proof-of-Stake (PoS) blockchain systems (unlike PoW) have a \emph{replicable} consensus resource: the executor (validator) stake can be invested in multiple blocks at the same time. Without risking their stake, executors in PoS systems can notarize (validate) conflicting blocks, thereby launching a \emph{nothing-at-stake} attack \cite{v-pos} and compromising the consistency of the blockchain. \\ 
\noindent Fortunately, in a regulated PoS blockchain, the regulated executors $R$ cannot mount a nothing-at-stake attack as they are mandated by the regulator to notarize blocks in the regulated branch, exclusively. However, there is a possibility that unregulated executors $\mybar{R}$ are the sole party that can do a costless simulation with a nothing-at-stake attack. This implies that the consensus resource invested on the regulated branch is $\geq \alpha_R$ and that on the unregulated branch is $\leq \alpha_{\mybar{R}}$, and all the results from our regulated blockchain stochastic games where the regulated branch weight has a known lower-bound on $\alpha_R$ (implying a lower-bound on the regulated branch weight in the PoS blockchain), directly apply. 
% Unlike Bitcoin, the process of consensus is truly distributed: notarization of a block does not depend on local hash power, but through validation votes from the peers of the block proposer. This allows all honest block validators to vote on blocks proposed by regulated executors. Consequently, in our theorems, $\alpha_R$ should be replaced with $\alpha_{R^*}$ (remembering the regulated executors $R$ are a subset of honest executors $R^*$), with a revised normalization: $\alpha_{R^*} + \alpha_{\mybar{R}} = 1$. The game depth is retained to be $E = \infty$, as presently, participating in Algorand's consensus protocol is unincentivized.

\subsection*{The Practical Implication of our Results}
\noindent We assume that the regulator $\mathcal{F}$ estimates $\alpha_R$ as part of its licensing practice, and accordingly conveys optimal consensus execution strategies to $R$. The blockchain network stabilizes once $\mybar{R}$ adopts best responses to the strategies chosen by $R$ (given by the Nash equilibria of our regulated blockchain stochastic games). \\
\noindent We are proposing a general regulatory framework for permissionless blockchains to give guarantees on legal block throughput as a function of the regulated consensus resource in the blockchain network. World governments can adopt this framework in general commercial cross-border cryptocurrencies like Bitcoin or Algorand, or setup localized digital currency networks in jurisdictionally constrained but permissionless `regulatory sandboxes'\footnote{A regulatory sandbox is a framework set up by a regulator to allow small scale and live testing of innovations by private firms in a controlled environment under the regulator's oversight \cite{reg-sandbox}.}.

%% file: 06_relatedwork.tex
\section{Related Work}
\label{sec:related}

We briefly review related work with respect to regulated asset trade using existing blockchain protocols designed for crypto-assets. Since none of the existing protocols have all the principles of regulation enforced within the consensus mechanism, we elucidate the legal and technical aspects of regulation enforced separately from the consensus mechanism. \\

\noindent \emph{Susceptibility of existing blockchains towards illegal transactions.} Regulators have identified that the illegal use of blockchain based virtual currencies includes money laundering and terror financing \cite{reg-gli-book}. There have been studies conducted for establishing emerging regulatory approaches for the future of blockchain based token economies \cite{reg-token}. None of the existing blockchain protocols, be them permissionless or permissioned, are designed with regulation in the consensus mechanism itself \cite{sok-xiao,presto,sok-bano}. This has introduced skepticism in the minds of regulatory authorities in the adoption of these protocols as is \cite{reg-promise}, and introduced regulation as a separate mechanism to be enforced by legal authorities. Even given separately enforced regulation, no comprehensive international regulatory framework exists for blockchain technology \cite{reg-trevor}, and cross jurisdictional governmental collaboration for regulated trade via legal transactions over blockchains is needed \cite{reg-peter}. Our regulatory framework achieves the said collaboration through support for cross-jurisdictional regulation. \\
\noindent \emph{Pocketed adoption of regulation on blockchains across the world.} Licensing for legal use of cryptocurrencies has been initiated by the New York financial services department, with the issuance of the `BitLicense' \cite{bitlicense} virtual currency business license. Many nations across the world, including U.K., Australia, the U.S., Hong Kong, Malaysia, Singapore, Switzerland, Thailand, and United Arab Emirates are either exploring or have implemented `regulatory sandboxes' for popular blockchains \cite{reg-nishith}. More specifically in Europe, federal banks \cite{reg-BoE} and governmental bodies \cite{reg-swiss} are contemplating (geographically constrained) oversight on existing blockchains. Unfortunately, these separate regulatory enforcements imply that cross-jurisdictional transaction conflict resolution would still require tenuous negotiations for (cross-jurisdictional) agreements on the price / commodity quantity being traded as part of the transaction. Such negotiations are trivial under our regulatory framework, owing to a cross-jurisdictional network of regulators overseeing the consensus on regulated asset trade. \\
\noindent \emph{Permissionless blockchains introduce an untrusted decentralized financial system in general.} Apart from jurisdiction, blockchain technology can be an enabler of decentralized autonomous organisations (DAOs), which have an uncertain legal status \cite{wiki-dao}, and so these DAOs constitute an untrusted financial system. These systems facilitate illegal transactions for which conflict resolution or penalisation claims cannot be legally enforced with complete authority \cite{reg-issues}. This drawback of blockchains has already resulted in a fallout for the acceptance of Bitcoin. \\
\indent Bitcoin has suffered a blow to being a trustworthy trade platform due to the large scale illegal trading in the Silk Road darknet market (as highlighted in Section \ref{sec:casestudy}). Given this illegal trade, and the existing mistrust in the deployment of Bitcoin for financial services in general, Bitcoin has been banned in many countries \cite{btc-legality}, and federal authorities in the US see more oversight coming on most cryptocurrencies in the future \cite{increased-oversight}. Our regulatory framework, on the other hand, provides a trustworthy, decentralized, cross-jurisdictional financial system by curtailing trader autonomy through regulation. \\
\noindent \emph{Consensus in State-of-the-art Permissionless Blockchains is insufficient.} Developers at Algorand admit \cite{reg-algorand} that developing a fast and secure permissionless blockchain is insufficient for a cross-jurisdictional economy, and there is a need to enforce regulation by leveraging sophisticated cryptographic constructions atop the consensus mechanism, such as verifiable random functions, Boneh-Lynn-Shacham aggregate signatures, and new Pixel multi-signatures \cite{reg-algorand}. These constructions introduce extraneous layers of computation apart from the consensus mechanism, and still are susceptible to attacks as long as their network is permissionless \cite{algorand-net}. Our regulatory framework does not need non-trivial cryptographic constructions, and thus is more suitable for regulated decentralized trade. \\
\noindent \emph{Our regulatory framework differs from federated blockchains.} Federated blockchains \cite{fedchain}, have a permissioned network where certain pre-selected nodes, from each organization that is maintaining a distributed ledger, have the authority to participate in the consensus protocol for agreeing on transactions. Unlike our regulatory framework, where a trader has the authority to participate in consensus under the oversight of the regulator, federated blockchains have no notion of regulatory intervention for oversight in the transaction agreement process. \\
% \noindent  \emph{Inability of existing blockchains to define a perfectly fair economy.} Permissioned blockchains are inherently centralized in their consensus protocol, and consequently the block proposal by the trusted nodes opens the economy to bias / manipulation. Permissionless blockchains, owing to their resource based consensus (where the resource could be hash power, or stake for instance) can also bias the transaction agreement towards the resource wealthy members of the network, implying the impossibility of full decentralization \cite{imposs-fulldecen}. Our regulated blockchain framework eliminates this impossibility through the regulator (as a trusted third party), by inducing block proposal which is fair for all transactors.

%% file: 07_conclusion.tex
\section{Conclusion and Future Directions}

\noindent In this contribution, we have proposed a framework for the legal trade of regulated assets via cryptocurrencies through the appropriate regulation of the consensus resource of existing blockchain consensus protocols. First, we have motivated the need for our framework through a case study of the Silkroad deepweb cryptocurrency based market. Next, we have formally presented our regulatory framework and have given guarantees of legal transaction throughput as a function of the total fraction of the regulated consensus resource. We have shown that the legal transaction throughput can be maximized when the regulated consensus resource is in a majority in the blockchain network, and the regulated protocol executors follow a `longest legal branch' block notarization rule. \\
\noindent In future, we would like to analyze competition between regulated and unregulated executors, when the unregulated executors play the \texttt{DubFrontier} strategy with a non-zero pay-forward scheme, but the regulated executors are not supported by the regulator through an oversight compliance fee. We would like to derive conditions for a successful long range attack by regulated validators on a vanilla proof-of-stake blockchain to confirm a legal branch originating from the genesis block. Finally, we would like to perform a formal, novel Markov Decision Process (MDP) analysis for block proposal competition between regulated and unregulated executors, in both proof-of-work and proof-of-stake regimes, similar to the recent proposal in \cite{sm-mdp}.

\section*{Acknowledgement}
We would really like to thank Leana Golubchik (Professor, USC) for her comments on improving the readability of a previous version of this paper.

% \section*{Contribution Statement}
% All authors jointly formulated the problem. Aditya Ahuja and Vinay J. Ribeiro designed the regulated blockchain system model. Aditya Ahuja conducted the case study, and constructed the regulated consensus protocols (both for legal and fully decentralized consensus). Ranjan Pal proposed the competition analysis between regulated and unregulated executors, and validated the blockchain stochastic game. Aditya Ahuja proposed the mathematical reduction of the executor competition to the blockchain stochastic game. Vinay J. Ribeiro and Leana Golubchik validated all the algorithmic constructions and competition analysis.

%% file: 09a_thm-proofs.tex
\newpage
\section{Theorem Proofs}
\label{app:thm-proofs}

\noindent We now give the construction of our regulated blockchain stochastic games, reduce their construction to existing blockchain mining games, and prove the theorems corresponding to the main results of our framework.

\subsection{Preliminaries}

\subsubsection*{Blockchain Game with Regulated Executor Dominance}
\noindent We define our first regulated blockchain stochastic game \textsf{RBChain-Rdom-SGame}, applicable when the regulated executors are in a majority in the blockchain network ($\alpha_R > 0.5$). Given a root block, the blockchain state is given by $(b_{\mybar{R}},b_R)$, where $b_{\mybar{R}}$ blocks are proposed by the unregulated executors, and $b_R$ blocks are proposed by the regulated executor. The regulator may include an OCF in all of $b_R$ blocks.

\noindent \fbox{\begin{minipage}{26em}
\textsf{RBChain-Rdom-SGame} States:
\begin{itemize}
\item \emph{Mining States.} This set of states, denoted by $M$, is a collection of states $(b_{\mybar{R}},b_R)$ where both $R$ and $\mybar{R}$ notarize blocks on their own branch. Note that $(0,0) \in M$.
\item \emph{Defection/Capitulation States.} This set of states, denoted by $C$, is when executor $\mybar{R}$ defects and abandons its own branch, and adds (legal or dubious) blocks linked to some regulated block in the competing regulated branch, transitioning the game from state $(b_{\mybar{R}},b_R)$ to state $(0,s_R)$ where $s_R \in \{ 0, 1, ..., b_R - 1 \}$.
\item \emph{Legal Winning States.} This set of states is given by $W_{leg} := \{ (b_{\mybar{R}},b_R) : b_{\mybar{R}} = b_R + 1 \}$ and all $b_{\mybar{R}}$ blocks in each state of $W_{leg}$ are legal. Under $W_{leg}$, when executor $\mybar{R}$ overtakes, the game transitions to state $(0,0)$.
\end{itemize}
\end{minipage}}

\subsubsection*{Blockchain Game with Unregulated Executor Dominance}
\noindent We define our second regulated blockchain stochastic game \textsf{RBChain-URdom-SGame}, applicable when the unregulated executors are in a majority in the blockchain network ($\alpha_{\mybar{R}} > 0.5$). Given a root block, the blockchain state is given by $(b_R,b_{\mybar{R}})$, where $b_{\mybar{R}}$ blocks are proposed by the unregulated executors, and $b_R$ blocks are proposed by the regulated executors. The regulated executors release $min(b_R,b_{\mybar{R}})$ blocks.

\noindent \fbox{\begin{minipage}{26em}
\textsf{RBChain-URdom-SGame} States:
\begin{itemize}
\item \emph{Mining States.} This set of states, denoted by $M$, is a collection of states $(b_R,b_{\mybar{R}})$ where both $\mybar{R}$ and $R$ notarize blocks on their own branch. Note that $(0,0) \in M$.
\item \emph{Dubious Cut Defection/Capitulation States.} This set of states, denoted by $C_{dub}$, is when executor $R$ defects and abandons its own branch, and adds regulated blocks linked to some regulated block in the competing dubious branch, transitioning the game from state $(b_R,b_{\mybar{R}})$ to state $(0,s_{\mybar{R}})$ where $s_{\mybar{R}} \in \{ 0, 1, ..., b_{\mybar{R}} - 1 \}$, and all $s_{\mybar{R}}$ blocks are dubious.
\item \emph{Winning States.} This set of states is given by $W := \{ (b_R,b_{\mybar{R}}) : b_R = b_{\mybar{R}} + 1 \}$. Under $W$, when executor $R$ overtakes, the game transitions to state $(0,0)$.
\end{itemize}
\end{minipage}}

\subsubsection*{Original Blockchain Stochastic Games}
\noindent We will refer to the original blockchain mining game under the immediate release model, proposed by Kiayias et.al. (Sections 2 and 3 in \cite{bmg}) as the \textsf{K1-IR-SGame}. We will refer to the blockchain mining game with a pay-forward scheme, under the immediate release model, proposed by Koutsoupias et.al. (Sections 2 and 3 in \cite{bmg-pf}) as the \textsf{K2-IR-SGame}. Also, we will refer to the original blockchain mining game under the strategic release model (Section 4 in \cite{bmg}) as the \textsf{K1-SR-SGame}. \\
Both blockchain mining games \cite{bmg,bmg-pf} do not consider a distinction between legal and dubious blocks, and define the \texttt{Frontier} mining strategy as a choice to mine on the deepest block in the blockchain among all competing branches (with or without a pay-forward). 

\subsection{Proof of Theorem 8}
\label{subapp:thm1}

\noindent The \textsf{K1-IR-SGame} considers two miners, named $1$ and $2$, where miner $2$ is in a majority in the Bitcoin network ($\alpha_2 > 0.5$), and is always following the \texttt{Frontier} strategy. Theorem 3.2 \cite{bmg} from \textsf{K1-IR-SGame} proves that when miner $1$ has hash power $\alpha_1$ less than the root of the polynomial $2\alpha^2 - (1 - \alpha)^3 \approx 0.361$, then \texttt{Frontier} is a Nash equilibrium strategy for miner $1$. Theorem 3.2 is proven by eliminating the possible set of mining states under the given condition. Next, through Theorem 3.12 \cite{bmg} in the \textsf{K1-IR-SGame}, it has been proven that, for game depth E = 3, the expected gain $g_1$ per epoch of miner $1$ is equal to $\frac{\alpha_1^2(2 + 2\alpha_1 - 5\alpha_1^2 + 2\alpha_1^3)}{1 - \alpha_1^2 + 2\alpha_1^3 - \alpha_1^4} > \alpha_1$, by demonstrating strategies more rewarding than the \texttt{Frontier} strategy for $\alpha_1 > 0.455$. By considering alternate game depths $E$, the \textsf{K1-IR-SGame} experimentally establishes (Table 1 in \cite{bmg}) the lower bound for a deviating strategy to be $\alpha_1 > h_{IR} \approx 0.42$. \\ 
\noindent The \textsf{RBChain-Rdom-SGame} reduces to an instance of the \textsf{K1-IR-SGame}, when the first miner is the unregulated executor $\mybar{R}$, the second miner is the regulated executor $R$, the first miner's \texttt{Frontier} strategy is replaced by the \texttt{LegFrontier} strategy, and the second miner's \texttt{Frontier} strategy is replaced by the \texttt{RegFrontier} strategy. Consequently, the results on the threshold $h_{IR}$ from the \textsf{K1-IR-SGame} directly apply to the \textsf{RBChain-Rdom-SGame}.

\subsection{Proof of Theorem 9}

\noindent The \textsf{K2-IR-SGame} has an identical setting to the \textsf{K1-IR-SGame} in terms of defining the miners, and their best response strategies. However, \textsf{K2-IR-SGame} allows miner $1$ to add a pay-forward reward $w$ (as some unknown function of $\alpha_1$) to the blocks mined by it. This reward $w$ is collected by the miner who confirms a block immediately succeeding the block that contains the announcement of $w$. In this setting, it is proven through Theorem 3.2 \cite{bmg-pf}, that \texttt{Frontier} is a Nash equilibrium strategy for miner $1$, when miner $2$ has consensus resource $\alpha_2 > h^{ocf}_{IR} = 0.5$. \\
\noindent The \textsf{RBChain-Rdom-SGame} with a regulator contributed OCF $\rho_{\mathcal{F}}$ in the regulated blocks, reduces to an instance of the \textsf{K2-IR-SGame}, when the first miner is the unregulated executor $\mybar{R}$, the second miner is the regulated executor $R$, the first miner's \texttt{Frontier} strategy is replaced by the \texttt{LegFrontier} strategy, and the second miner's \texttt{Frontier} strategy is replaced by the \texttt{RegFrontier}$(\rho_{\mathcal{F}})$ strategy. Consequently, the results on the threshold $h^{ocf}_{IR}$ from the \textsf{K2-IR-SGame} directly apply to the modified \textsf{RBChain-Rdom-SGame}.

\subsection{Proof of Theorem 10}

\noindent The \textsf{K1-SR-SGame} again considers two miners, named $1$ and $2$, where miner $2$ is in a majority in the Bitcoin network ($\alpha_2 > 0.5$), and is always following the \texttt{Frontier} strategy while immediately releasing blocks. However, when it comes to miner $1$, in the \textsf{K1-SR-SGame}, the said miner only releases $min(b_1,b_2)$ blocks, where $b_i$ blocks are successfully mined by miner $i \in \{ 1,2 \}$. In this regime, it is proven in Theorem 4.1 \cite{bmg}, that the best response for miner $1$ is \texttt{Frontier} only when $\alpha_1$ is less than the root of the polynomial $\alpha^3 - 6\alpha^2 + 5\alpha - 1 \approx 0.308$, but this bound can be improved to $0.33$ through results in \cite{btc-osm,bmg-pf}. \\
\noindent The \textsf{RBChain-URdom-SGame} reduces to an instance of the \textsf{K1-SR-SGame}, when the first miner is the regulated executor $R$ following strategic release, the second miner is the unregulated executor $\mybar{R}$, the first miner's \texttt{Frontier} strategy is replaced by the \texttt{RDubFrontier} strategy, and the second miner's \texttt{Frontier} strategy is replaced by the \texttt{DubFrontier} strategy. Consequently, the results on the threshold $\hat{h}_{SR}$ from the \textsf{K1-SR-SGame} directly apply to the \textsf{RBChain-URdom-SGame}.

%% file: RBChain-main.bbl
\begin{thebibliography}{10}

% Regulation (Principles)
\bibitem{reg-principles} Stiglitz, Joseph E. ``Government failure vs. market failure: Principles of regulation." (2008).

\bibitem{nn-regstate} Stocker, Volker, Georgios Smaragdakis, and William Lehr. ``The state of network neutrality regulation." ACM SIGCOMM Computer Communication Review 50, no. 1 (2020): 45-59.

\bibitem{capmarket-reg} The Indian Capital Market Regulators: Roles and Functions \\
\url{https://www.elearnmarkets.com/blog/capital-market-regulators/}

% Regulation and Blockchains (Legal/Business Studies/Articles)
\bibitem{reg-peter} Yeoh, Peter. ``Regulatory issues in blockchain technology." Journal of Financial Regulation and Compliance (2017).
\bibitem{reg-trevor} Kiviat, Trevor I. ``Beyond bitcoin: Issues in regulating blockchain transactions." Duke LJ 65 (2015): 569.
\bibitem{reg-dlt} Bahlke, Conrad G., and Marija Pecar. "Unblocking the Blockchain: Regulating Distributed Ledger Technology." The Journal on the Law of Investment and Risk Management Products 36, no. 10 (2016).
\bibitem{reg-decen} Nabilou, Hossein. ``How to regulate bitcoin? Decentralized regulation for a decentralized cryptocurrency." International Journal of Law and Information Technology 27, no. 3 (2019): 266-291.
\bibitem{bitlicense} Chohan, Usman W. "Oversight and regulation of cryptocurrencies: BitLicense." Available at SSRN 3133342 (2018).
\bibitem{reg-gli-book} Global Legal Insights - Blockchain and Cryptocurrency Regulation 2019, First Edition (Online; Accessed 2020-05-21) \\ \url{https://www.acc.com/sites/default/files/ \\ resources/vl/membersonly/Article/1489775-1.pdf}
\bibitem{reg-nishith}
Nishith Desai Associates - The Blockchain : Industry Applications and Legal Perspectives (Online; Accessed 2020-05-21) \\ \url{http://www.nishithdesai.com/fileadmin/user-upload/ \\ pdfs/Research-Papers/The-Blockchain.pdf}
\bibitem{reg-promise}
The Conversation - People don't trust blockchain systems - Is regulation a way to help? (Online; Accessed 2020-05-21) \\ \url{https://theconversation.com/people-dont-trust- \\ blockchain-systems-is-regulation-a-way-to-help-110007}
\bibitem{reg-sandbox}
Jenik, Ivo, and Kate Lauer. ``Regulatory sandboxes and financial inclusion." Washington, DC: CGAP (2017).
\bibitem{reg-swiss}
Reuters - Swiss task force to look into blockchain oversight (Online; Accessed 2020-05-21) \\ \url{https://www.reuters.com/article/us-swiss-crypto/ \\ swiss-task-force-to-look-into-blockchain-oversight-idUSKBN1F71T6}
\bibitem{reg-BoE}
CoinDesk - Bank of England Eyes Private Blockchain Oversight [Online; Accessed 2020-05-21] \\ \url{https://www.coindesk.com/bank-of-england-eyes- \\ regulatory-oversight-of-private-blockchain-data}
\bibitem{wiki-dao}
Wikipedia - Decentralized Autonomous Organisations (Online; Accessed 2020-05-21) \\ \url{https://en.wikipedia.org/wiki/ \\ Decentralized-autonomous-organization}
\bibitem{reg-issues} Blocks Decoded - 5 Blockchain Problems: Security, Privacy, Legal, Regulatory, and Ethical Issues (Online; Accessed 2020-05-21) \\ \url{https://blocksdecoded.com/blockchain-issues- \\ security-privacy-legal-regulatory-ethical/} 
\bibitem{btc-legality}
Wikipedia - Legality of bitcoin by country or territory [Online; Accessed 2020-05-21] \\ \url{https://en.wikipedia.org/wiki/ \\ Legality-of-bitcoin-by-country-or-territory}
\bibitem{increased-oversight}
Legal Executive Institute - Consensus 2019: Blockchains Biggest US Conference Sees Increased Legal Oversight Coming (Online; Accessed 2020-05-21) \\ \url{http://www.legalexecutiveinstitute.com/consensus- \\ 2019-blockchain/}
\bibitem{reg-algorand}
Algorand: Where Cryptography meets Economics (Online; Accessed 2020-05-21) \\ \url{https://www.algorand.com/what-we-do/economic- \\ innovation/where-cryptography-meets-economics/}
\bibitem{fedchain} 
2019 The Year of the Federated Blockchain ? Blockchain Consortium Simply Explained (Online; Accessed 2020-05-21) \\ \url{https://101blockchains.com/federated-blockchain/}
\bibitem{reg-token} Ferreira, Agata. "Emerging Regulatory Approaches to Blockchain Based Token Economy." The Journal of The British Blockchain Association (2020): 12270.
\bibitem{reg-mon} Peters, Gareth, Efstathios Panayi, and Ariane Chapelle. ``Trends in cryptocurrencies and blockchain technologies: A monetary theory and regulation perspective." Journal of Financial Perspectives 3, no. 3 (2015).
\bibitem{ifc-bchain} International Finance Corporation - Blockchain Opportunities for Private Enterprises in Emerging Markets (Online; Accessed 2020-05-21) \\ \url{https://www.ifc.org/wps/wcm/connect/2106d1c6-5361-41cd-86c2-f7d16c510e9f/201901-IFC- EMCompass-Blockchain-Report.pdf-MOD=AJPERES-CVID=mxYj-sA}

\bibitem{btcnews-yellen} Bitcoin News - Janet Yellen Reveals Plans for Bitcoin, Sees Cryptocurrencies Used Mainly for Illicit Financing (Online; Accessed 2021-01-26) \\ \url{https://news.bitcoin.com/janet-yellen-bitcoin-cryptocurrencies-illicit-financing/}

\bibitem{btcnews-lagarde} Bitcoin News - ECB Chief Christine Lagarde Calls for Global Bitcoin Regulation, Says BTC Conducts `Funny Business' (Online; Accessed 2021-01-26) \\ \url{https://news.bitcoin.com/ecb-christine-lagarde-global-bitcoin-regulation-btc/}


% Blockchain (Consensus)
\bibitem{sok-bano} Bano, Shehar, Alberto Sonnino, Mustafa Al-Bassam, Sarah Azouvi, Patrick McCorry, Sarah Meiklejohn, and George Danezis. ``SoK: Consensus in the age of blockchains." In Proceedings of the 1st ACM Conference on Advances in Financial Technologies, pp. 183-198. 2019.

\bibitem{sok-wang} Wang, Wenbo, Dinh Thai Hoang, Peizhao Hu, Zehui Xiong, Dusit Niyato, Ping Wang, Yonggang Wen, and Dong In Kim. ``A survey on consensus mechanisms and mining strategy management in blockchain networks." IEEE Access 7 (2019): 22328-22370.

\bibitem{sok-xiao} Xiao, Yang, Ning Zhang, Wenjing Lou, and Y. Thomas Hou. ``A survey of distributed consensus protocols for blockchain networks." IEEE Communications Surveys and Tutorials 22, no. 2 (2020): 1432-1465.

\bibitem{presto} Leonardos, Stefanos, Daniël Reijsbergen, and Georgios Piliouras. ``PREStO: A systematic framework for blockchain consensus protocols." IEEE Transactions on Engineering Management (2020).

\bibitem{bitcoin} Nakamoto, Satoshi. ``Bitcoin: A peer-to-peer electronic cash system". Manubot, 2019.

\bibitem{v-pos} Li, Wenting, Sébastien Andreina, Jens-Matthias Bohli, and Ghassan Karame. "Securing proof-of-stake blockchain protocols." In Data Privacy Management, Cryptocurrencies and Blockchain Technology, pp. 297-315. Springer, Cham, 2017.

\bibitem{algorand} Gilad, Yossi, Rotem Hemo, Silvio Micali, Georgios Vlachos, and Nickolai Zeldovich. ``Algorand: Scaling byzantine agreements for cryptocurrencies." In Proceedings of the 26th Symposium on Operating Systems Principles, pp. 51-68. 2017.
\bibitem{algorand-net}
Algorand: A Permissionless Blockchain (Online; Accessed 2020-05-21) \\ \url{https://www.algorand.com/what-we-do/technology/ \\ permissionless-blockchain/}
\bibitem{algorand-ppos}
Algorand: A Pure Proof-of-Stake Approach (Online; Accessed 2020-05-21) \\ \url{https://www.algorand.com/what-we-do/technology/pure-proof-of-stake/}

% \bibitem{rapidchain} Zamani, Mahdi, Mahnush Movahedi, and Mariana Raykova. ``Rapidchain: Scaling blockchain via full sharding." In Proceedings of the 2018 ACM SIGSAC Conference on Computer and Communications Security, pp. 931-948. 2018.

% \bibitem{pala} Chan, T-H. Hubert, Rafael Pass, and Elaine Shi. ``PaLa: A Simple Partially Synchronous Blockchain." IACR Cryptol. ePrint Arch. 2018 (2018): 981.

\bibitem{streamlet} Chan, Benjamin Y., and Elaine Shi. "Streamlet: Textbook streamlined blockchains." In Proceedings of the 2nd ACM Conference on Advances in Financial Technologies, pp. 1-11. 2020.

% \bibitem{imposs-fulldecen} Kwon, Yujin, Jian Liu, Minjeong Kim, Dawn Song, and Yongdae Kim. ``Impossibility of full decentralization in permissionless blockchains." In Proceedings of the 1st ACM Conference on Advances in Financial Technologies, pp. 110-123. 2019.

% \bibitem{tensorflip} Ahuja, Aditya. ``TensorFlip: A Fast Fully-Decentralized Computational Lottery for Cryptocurrency Networks." To Appear In the 13th International Conference on Communication Systems and Networks (COMSNETS). IEEE, 2021.


% Blockchain (Regulation and Competition)
\bibitem{chain-reg-anon} Li, Yannan, Willy Susilo, Guomin Yang, Yong Yu, Xiaojiang Du, Dongxi Liu, and Nadra Guizani. ``Toward privacy and regulation in blockchain-based cryptocurrencies." IEEE Network 33, no. 5 (2019): 111-117.

\bibitem{btc-sm1} Eyal, Ittay, and Emin Gun Sirer. "Majority is not enough: Bitcoin mining is vulnerable." In International conference on financial cryptography and data security, pp. 436-454. Springer, Berlin, Heidelberg, 2014.

\bibitem{btc-osm} Sapirshtein, Ayelet, Yonatan Sompolinsky, and Aviv Zohar. ``Optimal selfish mining strategies in bitcoin." In International Conference on Financial Cryptography and Data Security, pp. 515-532. Springer, Berlin, Heidelberg, 2016.

\bibitem{bmg} Kiayias, Aggelos, Elias Koutsoupias, Maria Kyropoulou, and Yiannis Tselekounis. ``Blockchain mining games." In Proceedings of the 2016 ACM Conference on Economics and Computation, pp. 365-382. 2016.

\bibitem{bmg-pf} Koutsoupias, Elias, Philip Lazos, Foluso Ogunlana, and Paolo Serafino. ``Blockchain mining games with pay forward." In The World Wide Web Conference, pp. 917-927. 2019.

\bibitem{btc-minepractice} Khairuddin, Irni Eliana, and Corina Sas. ``An Exploration of Bitcoin mining practices: Miners' trust challenges and motivations." In Proceedings of the 2019 CHI Conference on Human Factors in Computing Systems, pp. 1-13. 2019.


% Cryptography and Game Theory
\bibitem{crypto-kl} Katz, Jonathan, and Yehuda Lindell. ``Introduction to modern cryptography". CRC press, 2014.
\bibitem{stochastic-games} Shapley, Lloyd S. ``Stochastic games." Proceedings of the national academy of sciences 39, no. 10 (1953): 1095-1100.
\bibitem{sg-solan} Solan, Eilon, and Nicolas Vieille. ``Stochastic games." Proceedings of the National Academy of Sciences 112, no. 45 (2015): 13743-13746.

% Silk Road
\bibitem{silk-econ} Hardy, Robert Augustus, and Julia R. Norgaard. ``Reputation in the Internet black market: an empirical and theoretical analysis of the Deep Web." Journal of Institutional Economics 12, no. 3 (2016): 515-539.

\bibitem{silk-drug} Van Hout, Marie Claire, and Tim Bingham. ``Surfing the Silk Road: A study of users' experiences." International Journal of Drug Policy 24, no. 6 (2013): 524-529.

\bibitem{silk-law} Trautman, Lawrence J. ``Virtual currencies; Bitcoin and what now after Liberty Reserve, Silk Road, and Mt. Gox?." Richmond Journal of Law and Technology 20, no. 4 (2014).

\bibitem{silk-bust} CNBC - Silk Road Cryptocurrency Bust by the US Government (Online; Accessed 06-Nov-2020)  \\
\url{https://www.cnbc.com/2020/11/05/1-billion-worth-of-bitcoin-linked \\ -to-the-silk-road-seized-by-the-us.html}

\bibitem{silk-beyond} Kleiman, Jared A. ``Beyond the silk road: unregulated decentralized virtual currencies continue to endanger US national security and welfare." Nat'l Sec. L. Brief 4 (2013): 59.

\bibitem{silk-lost} Martin, James. ``Lost on the Silk Road: Online drug distribution and the `cryptomarket'." Criminology and Criminal Justice 14, no. 3 (2014): 351-367.

% State-of-the-World Articles
\bibitem{cc-mcap} Reuters - Crypto market cap surges above \$1 trillion for first time (Online; Accessed 2021-02-14) \\
\url{https://www.reuters.com/article/crypto-currency-int-idUSKBN29C264}

\bibitem{evd-ccreg} Regulation of Cryptocurrency Around the World (Online; Accessed 2021-02-14) \\
\url{https://www.loc.gov/law/help/cryptocurrency/world-survey.php}

%\bibitem{evd-illegal} Evidence of illegal activities atop Blockchain based Cryptocurrencies: \\
%\url{https://www.elliptic.co/our-thinking/bitcoin-money-laundering} \\
%\url{https://www.investopedia.com/terms/b/blockchain.asp} \\
%\url{https://academic.oup.com/rfs/article/32/5/1798/5427781} \\
%\url{https://www.forbes.com/sites/rachelwolfson/2018/11/26/tracing-illegal-activity-through-the-bitcoin-blockchain-to\\-combat-cryptocurrency-related-crimes/\#6c91bdcf33a9} \\
%\url{https://www.nytimes.com/2020/01/28/technology/bitcoin-black-market.html}

\bibitem{evd-status} Evidence of non-determinate legal status of Blockchain based Cryptocurrencies: \\
\url{https://www.businessinsider.in/finance/news/how-the-laws-regulations-affecting-blockchain-technology-and-\\cryptocurrencies-like-bitcoin-can-impact-its-adoption/articleshow/74464680.cms}

% Future Directions
\bibitem{sm-mdp} Zur, Roi Bar, Ittay Eyal, and Aviv Tamar. ``Efficient MDP analysis for selfish-mining in blockchains." In Proceedings of the 2nd ACM Conference on Advances in Financial Technologies, pp. 113-131. 2020.

\end{thebibliography}
